\documentclass[aps,twocolumn,prb,amsmath,amssymb,showpacs]{revtex4}
\usepackage{multirow}
\usepackage{graphicx,epsfig}
\usepackage{wasysym}
\usepackage{float}

\begin{document}
\title{Density of states of interacting quantum wires with impurities:\\a Dyson equation approach}
\author{R. Zamoum$^{1}$}
\author{M. Guigou$^{2,3}$}
\author{C. Bena$^{2,3}$}
\author{A. Cr\'epieux$^{1}$}
\affiliation{$^1$Aix Marseille Universit\'e, Universit\'e de Toulon, CNRS, CPT UMR 7332, 13288 Marseille, France}
\affiliation{$^2$Institut de Physique Th\'eorique, CEA/Saclay, Orme des Merisiers, 91190 Gif-sur-Yvette Cedex, France}
\affiliation{$^3$Laboratoire de Physique des Solides, UMR 8502, B\^at. 510, 91405 0rsay Cedex, France}

\begin{abstract}
We calculate the density of states for an interacting quantum wire in
the presence of two impurities of arbitrary potential strength. To
perform this calculation, we describe the Coulomb interactions in the
wire within the Tomonaga-Luttinger liquid theory. After establishing
and solving the Dyson equation for the fermionic retarded Green's
functions, we study how the profile of the local density of states is
affected by the interactions in the entire range of impurity
potentials. Same as in the non-interacting case, when increasing the
impurity strength, the central part of the wire becomes more and more
disconnected from the semi-infinite leads, and discrete localized
states begin to form; the width and the periodicity of the corresponding peaks in the
spectrum depends on the interaction strength. As expected from the
Luttinger liquid theory, impurities also induce a reduction of the
local density of states at small energies. Two other important aspects
are highlighted: the appearance of an extra modulation in the density
of states at nonzero Fermi momentum when interactions are present,
and the fact that forward scattering must be taken into account in
order to recover the Coulomb-blockade regime for strong impurities.
\end{abstract}

\maketitle

\section{Introduction}

The interplay between interactions and disorder is a long-standing problem in condensed matter physics. In one-dimensional systems, in which the interactions can be treated exactly using the Luttinger liquid theory\cite{tomonaga50,luttinger63,mattis63} and bosonization,\cite{haldane81,gogolin98} a lot of progress to understand the effects of impurities has been made over the last 20 years. It was shown that repulsive interactions such as the Coulomb interactions renormalize the impurity strength, such that at low energy even a weak impurity has a very strong effect and can cut the wire into two pieces.\cite{glazman92,kane92a,kane92b,furusaki93,fisher97} This translates into a reduction of the local density of states (LDOS) at low energies, and the LDOS decays to zero as a power law.\cite{maslov95,fabrizio97,egger98,dolcini03,dolcini05} At high energies, the effect of the impurity consists in a small power-law correction of the unperturbed LDOS. The two power laws are characterized by two different exponents which depend on the interaction strength. 

These two regimes have been described perturbatively in the framework of the Luttinger liquid model by various techniques, e.g., the renormalization group,\cite{andergassen04,metzner12} the Keldysh formalism,\cite{dolcini03,dolcini05} or the duality between the weak impurity regime and the strong impurity regime.\cite{weiss96} These techniques have as starting point either the infinite clean wire, or a system of two decoupled semi-infinite wires and, next, one considers small impurity-induced perturbations around these points. However, the transition between the two limits cannot be captured by perturbative techniques. Then, special techniques such as the Bethe ansatz,\cite{saleur95} refermionization methods,\cite{kane92a,weiss95} approaches based on resummation of terms in perturbation theory,\cite{yue94,nazarov03,polyakov03,aristov09} or non perturbative-in-tunneling fixed-point method\cite{aristov10} are necessary. It would be thus of great interest to be able to capture this transition by more direct methods such as the Dyson equation technique we propose here. This is our first motivation to tackle this problem.


Our second motivation comes from a more applied perspective, and consists in providing a method to disentangle the effects of the metallic contacts which are inevitably connected to an interacting quantum wire (QW) in the measurement process, allowing one to have access to the interacting physics in the wire and eventually evaluate the strength of the interactions therein. The first attempts to measure the interaction parameters in a QW rely on the existence of the impurity-induced power-law corrections in the LDOS detectable in conductance measurements.\cite{fisher97,liang01} However, this has turned out to be a very difficult task, due especially to the incertitudes in fitting power-law dependencies over small intervals of energies \cite{}. Subsequently, other more direct measurements such as the shot noise have been proposed.\cite{bena01,trauzettel04,safi08,guigou07,recher06} Unfortunately, it was shown that in such experiments, the metallic contacts prevent one from having access to the value of the interacting parameter.\cite{safi95,trauzettel04,dolcini05,herrmann07,kim07}

The modelization of the metallic contacts provides us with two challenges. The first consists in the introduction of two impurities at the two junctions between the wire and the contacts. In general, the impurity potentials induced by these impurities are neither too small nor too large, and a non perturbative technique would be required to capture the transition between small and large values of the impurity potential. Secondly, the physics of the system is greatly affected by the fact that the two semi-infinite metallic leads are non interacting, and thus a correct modelization of the system needs to include the spatial inhomogeneity in the interaction strength. As mentioned above, this inhomogeneity is what blocks one from having access to the value of the interacting parameter via shot noise experiments. These two aspects need to be taken into account to correctly evaluate the effect of the contacts.

The first issue, i.e., the transition between the small and large impurities is an interesting problem which is present even in the absence of interactions. For a non-interacting system it has been shown that a transition from a Fabry-Perot to a Coulomb-blockade regime occurs when the impurity strength increases. Thus, for weak impurities, Fabry-Perot oscillations arise in the dependence of the LDOS on energy.\cite{bockrath99,yao99,kim07,wu07} For large values of the impurity potential, the system is cut into a central quantum-dot-like region plus two semi-infinite wires, and the energy spectrum of the central part consists in discrete levels whose energy is proportional to its inverse length. \cite{postma01,thorwart02,kleinmann02,thorwart05,tans97,bockrath97,nygard99,liang02,rubio00} The width of these states becomes smaller and smaller when the central part is more and more disconnected from the leads. 

Understanding the effect of the interactions on the transition between the Coulomb-blockade and the Fabry-Perot regime is a long-standing mesoscopic physics problem. Various other factors also come into place, such as lifting the degeneracy between the energy levels when Coulomb interactions are present. This is especially interesting in the spinful case: in this situation, the periodicity of oscillations in the LDOS is expected to change between the Coulomb-blockade and the Fabry-Perot  regimes; various works have been trying to approach this problems using different methods.\cite{peca03,alicea04}

In this work, we develop an approach that allows one to study the interplay between impurities and  interactions for arbitrary size impurities. Our approach is based on writing and solving the Dyson equations for a wire with one or two impurities. In this paper, we consider an infinite homogeneous interacting QW and the corresponding form for the fermionic Green's functions. This allows us to study the first aspect of the problem raised by the presence of the contacts, i.e., the presence of one or two impurities of arbitrary size.

We start with the study of an infinite homogeneous interacting QW with a single impurity. We calculate the form of the Friedel oscillations as well as the dependence of the LDOS with energy. For weak impurities we retrieve the expected Luttinger liquid power-law dependence of the impurity with energy at both low and high energies. For strong impurities at either high energies or large distances to impurity, we recover a power-law dependence with the same exponent $(K+K^{-1}-2)/2$ as for the weak-impurity regime, consistent with the Luttinger liquid predictions. However, at low energy and small distance, our approach fails to recover the transition  to a different power-law exponent characteristic to breaking the wire into two independence pieces, i.e., with exponent ($K^{-1}-1$). This comes from a drawback in the approximation used in our approach, which consists to neglect terms mixing impurity potentials and Coulomb interactions in the Dyson equation. This nevertheless does not affect the behavior at large distances/energies, and the validity of the main results of this paper, i.e., the dependence of the LDOS in a wire with two impurities and the transition from the weak-impurity regime to the strong-impurity regime.




For a homogeneous interacting wire with two impurities, we find that the main effect of interactions is to modify the amplitude of the Fabry-Perot oscillations, as well as the height and width of the Coulomb-blockade peaks. More precisely, when interactions are taken into account, in the weak-impurity limit, the oscillations are reduced,  while in the strong-impurity limit the peaks get wider and smaller. The interactions also affect the periodicity of the oscillations and the distance between the peaks. Moreover, same as for a single impurity, power-law dependencies of the LDOS with energy arise, and the LDOS is reduced to zero on the impurity sites.

Another confirmation of the validity of our approach at large distance and high energy is the agreement between the form that we obtain for the LDOS in a wire with two impurities, and that obtained via a completely different technique by Anfuso and Eggert in Ref.~\onlinecite{anfuso03} for a Luttinger box. Same as Ref.~\onlinecite{anfuso03}, we find that at non zero Fermi momentum there is an extra modulation in the space dependence of the LDOS which arises solely in the presence of interactions. 

The paper is organized as follows: We present the model in Sec. II, and the general solution to the Dyson equations for an arbitrary chiral wire with one or two impurities in Sec. III. In Sec. IV, we study the simple situation of a wire with  a single impurity. In Sec. V, we discuss the results for an infinite homogeneous wire with two impurities. We conclude in Sec. VI.


\section{Model}

\begin{figure}[H]
\begin{center}
\includegraphics[width=8cm]{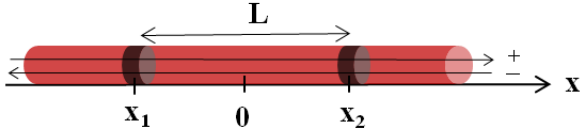}
\caption{One-dimensional wire with two impurities located at positions $x_{1,2}=\pm L/2$. The right ($+$) and left ($-$) chiralities are denoted by right and left arrows.}
\label{figure1}
\end{center} 
\end{figure}

We consider a one-channel interacting QW with two impurities at positions $x_{1,2}=\pm L/2$ (see Fig.~\ref{figure1}), where $L$ is the distance between the impurities. The impurities are described by backward scattering potentials $\lambda_{1,2}^\mathrm{B}$ and forward scattering potentials $\lambda_{1,2}^\mathrm{F}$. The Hamiltonian can be written as $H=H_0+H_\mathrm{int}+H_\mathrm{imp}$, where $H_0$ describes the non-interacting QW without impurities:
\begin{eqnarray}\label{H0}
H_0=-i\hbar v_F\sum_{r=\pm}r\int_{-\infty}^{\infty}\psi_{r}^{\dagger}(x)\partial_x\psi_{r}(x) dx~,
\end{eqnarray}
with $v_F$ is the Fermi velocity, $\psi_{r}^{\dagger}$ and $\psi_{r}$ are the creation and annihilation fermionic operators associated to the right movers ($r=+$) and left movers ($r=-$).  The Hamiltonian $H_\mathrm{int}$ describes the Coulomb interaction in the wire:
\begin{eqnarray}\label{Hint}
H_\mathrm{int}=\frac{1}{2}\int_{-\infty}^{\infty}\int_{-\infty}^{\infty}\hat\rho(x)V(x,x')\hat\rho(x')dxdx'~,
\end{eqnarray}
where $\hat\rho(x)=\sum_{r,r'}\psi_{r}^{\dagger}(x)\psi_{r'}(x)$ is the density operator, and $V$ is the Coulomb potential which is assumed to be short range due to screening effects by metallic gates or inter-wire coupling. The impurity Hamiltonian contains two types of contribution $H_\mathrm{imp}=H_B+H_F$, backward-scattering term
\begin{eqnarray}\label{HB}
H_{\mathrm{B}}=\sum_{r=\pm}\sum_{i=1,2}\int_{-\infty}^{\infty} \lambda_{i}^\mathrm{B}(x)\psi_{r}^{\dagger}(x)\psi_{-r}(x) dx~,
\end{eqnarray}
and forward-scattering term
\begin{eqnarray}\label{HF}
H_{\mathrm{F}}=\sum_{r=\pm}\sum_{i=1,2}\int_{-\infty}^{\infty} \lambda_{i}^\mathrm{F}(x)\psi_{r}^{\dagger}(x)\psi_{r}(x) dx~.
\end{eqnarray}
In the following, we assume that the impurities are localized, i.e. $\lambda_{i}^{\mathrm{F,B}}(x)=\Gamma_{i}^{\mathrm{F,B}}\delta(x-x_{i})$. Notice that in some works dealing with impurities in a Luttinger liquid, the forward-scattering terms were not included with the justification that such terms could be incorporated in the kinetic part.\cite{oreg96,furusaki97,vondelft98,grishin04} This is correct in the weak-impurity limit but does not hold in the
strong-impurity limit. Indeed, the density of states is strongly
affected by the forward-scattering terms at strong
$\Gamma_{1,2}^{\mathrm{F,B}}$; in particular, these terms need to be taken into
account explicitly in order to recover the Coulomb-blockade regime.

\section{Density of states of a wire with two impurities: General form}

The position-dependent density of states can be obtained from the generalized retarded Green's function as follows:
\begin{eqnarray}\label{dos}
\rho(x,\omega)=-\frac{1}{\pi}\sum_{r,r'}\mathrm{Im}\{G^{\mathrm{R}}_{r,r'}(x,x;\omega)\}~,
\end{eqnarray}
where the retarded Green's function $G^{\mathrm{R}}_{r,r'}(x,x';\omega)$ is defined as the Fourier transform of:
\begin{eqnarray}\label{G_R}
G^{\mathrm{R}}_{r,r'}(x,x';t,t')=-i\Theta(t-t')\langle\{\psi_{r}(x,t);\psi_{r'}^{\dagger}(x',t')\}\rangle~,
\end{eqnarray}
where $\Theta$ is the Heaviside function, and $\{a;b\}$ refers to the anticommutator.

In order to calculate the form of such Green's functions in the presence of impurities, we establish the Dyson equation associated to the Hamiltonian $H$. The details of the calculation are presented in Appendix \ref{dyson}. Assuming that Coulomb interactions are strongly attenuated with distance and weak in comparison to the energy and neglecting the contributions mixing the impurity potentials and the Coulomb potential, i.e., assuming that $|(x-x_i)\omega|\gg v_F$ (see Appendix \ref{approximation} for the details), we obtain:
\begin{eqnarray}\label{eq_dyson}
&&G_{r,r'}^\mathrm{R}(x,x';\omega)= g_{r}^\mathrm{R}\left(x,x';\omega\right)\delta_{r,r'}+\sum_{i=1,2} g_{r}^{\mathrm{R}}\left(x,x_i;\omega\right)\nonumber\\
&&\times\left[\Gamma_{i}^\mathrm{B} G_{-r,r'}^{\mathrm{R}}\left(x_i,x';\omega\right)
+\Gamma_{i}^\mathrm{F} G_{r,r'}^{\mathrm{R}}\left(x_i,x';\omega\right)\right]~,
\end{eqnarray} 
where $g_{r}^{\mathrm{R}}$ are the Green's functions of a clean interacting homogeneous wire, associated to $H_0+H_\mathrm{int}$. They can be obtained in the framework of the Tomonaga-Luttinger theory.\cite{tomonaga50,luttinger63} For an infinite QW with uniform interactions, their form has been derived explicitly in Ref.~\onlinecite{braunecker12}. These Green's functions depend on a single chiral index $r$ since the chiral states are eigenstates of the interacting QW in the absence of impurities:
\begin{eqnarray}\label{gfree}
&&g_{r}^{\mathrm{R}}(x,x';\omega)=\frac{-e^{irk_{F}(x-x')}K^2\omega_+}{2\hbar v_{F}\omega_c\sqrt{\pi}\;{\bf \Gamma}(1+\gamma)}\nonumber\\
&&\times \left( \frac{2i\vert x-x' \vert \omega_c}{aK\omega_{+}} \right)^{\frac{1}{2}-\gamma} \left[ {\bf K}_{\gamma-\frac{1}{2}}\left(\frac{K\vert x-x'\vert \omega_{+}}{ia\omega_c}\right)\right.\nonumber\\
&&-\left.\mathrm{sgn}(r(x-x')){\bf K}_{\gamma+\frac{1}{2}}\left(\frac{K\vert x-x' \vert \omega_{+}}{ia\omega_c}\right)\right]~,
\end{eqnarray}
where $\omega_+=\omega+i0$, $k_F$ is the Fermi momentum, $\omega_c=v_F/a$, $a$ is the small-distance cutoff of the Tomonaga-Luttinger liquid theory,\cite{tomonaga50,luttinger63} ${\bf \Gamma}$ and ${\bf K}$ are, respectively, the gamma and modified gamma functions, and $\gamma=(K+K^{-1}-2)/4$. Here, $K$ is the interaction parameter which is related to the interaction potential by the relation $K=[1+4V(k\approx 0)/(\pi v_F)]^{-1/2}$. In the non interacting limit, $K=1$, Eq.~(\ref{gfree}) recovers the chiral Green's functions of a clean non-interacting QW:
\begin{eqnarray}\label{gnon}
g_{r}^{\mathrm{R}}(x,x';\omega)&=&-\frac{ie^{irk_{F}(x-x')}}{\hbar v_F}\nonumber\\
&&\times e^{i\omega r(x-x')/v_F}\Theta(r(x-x'))~.
\end{eqnarray}

From Eq.~(\ref{eq_dyson}), we can extract the expressions of $G_{r,r}^{\mathrm{R}}\left(x_i,x';\omega\right)$ and $G_{-r,r}^{\mathrm{R}}\left(x_i,x';\omega\right)$ by solving a linear set of equations. We obtain (see Appendix \ref{solution} for the details of the calculation):
\begin{eqnarray}\label{Gxx1}
G_{r,r}^{\mathrm{R}}\left(x_i,x';\omega\right)&=&\frac{(1-\chi_r^{\bar{i}\bar{i}})g_r^{\mathrm{R}}(x_i,x';\omega)+\chi_r^{i\bar{i}}g_r^{\mathrm{R}}(x_{\bar{i}},x';\omega)}{(1-\chi_r^{11})(1-\chi_r^{22})-\chi_r^{12}\chi_r^{21}}~,\nonumber\\
\end{eqnarray}
and,
\begin{eqnarray}\label{Gxx2}
&&G_{-r,r}^{\mathrm{R}}(x_i,x';\omega)=\nonumber\\
&&D^{-1}\sum_{j=1,2}g_{-r}^{\mathrm{R}}\left(x_i,x_j;\omega\right)\Gamma_{j}^\mathrm{B}G_{r,r}^{\mathrm{R}}(x_j,x';\omega)\nonumber\\
&&+D^{-1}\Gamma_i^\mathrm{B}\Gamma_{\bar{i}}^\mathrm{F}\Big[g_{-r}^{\mathrm{R}}(x_i,x_{\bar{i}};\omega)g_{-r}^{\mathrm{R}}(x_{\bar{i}},x_i;\omega)\nonumber\\
&&-g_{-r}^{\mathrm{R}}(x_i,x_i;\omega)g_{-r}^{\mathrm{R}}(x_{\bar{i}},x_{\bar{i}};\omega)\Big]G_{r,r}^{\mathrm{R}}(x_i,x';\omega)~,
\end{eqnarray} 
with $\bar{i}=1$ when $i=2$, and $\bar{i}=2$ when $i=1$. Moreover, we have defined the quantities:
\begin{eqnarray}\label{def_D}
D&=&\left[1-\Gamma^\mathrm{F}_1g_{-r}^{\mathrm{R}}(x_1,x_1;\omega)\right]\left[1-\Gamma^\mathrm{F}_2g_{-r}^{\mathrm{R}}(x_2,x_2;\omega)\right]\nonumber\\
&&-\Gamma^\mathrm{F}_1g_{-r}^{\mathrm{R}}(x_1,x_2;\omega)\Gamma^\mathrm{F}_2g_{-r}^{\mathrm{R}}(x_2,x_1;\omega)~,
\end{eqnarray}
and,
\begin{eqnarray}\label{def_chi}
&&\chi_r^{ij}=\Gamma_i^\mathrm{F}g_{r}^{\mathrm{R}}(x_i,x_j;\omega)\nonumber\\
&&+D^{-1}\bigg[\Gamma_1^\mathrm{B}g_{r}^{\mathrm{R}}(x_i,x_{\bar j};\omega)\Gamma_2^\mathrm{B}g_{-r}^{\mathrm{R}}(x_{\bar i},x_i;\omega)\nonumber\\
&&+\Gamma_j^\mathrm{B}g_{r}^{\mathrm{R}}(x_i,x_j;\omega)\Gamma_{\bar j}^\mathrm{F}g_{-r}^{\mathrm{R}}(x_1,x_2;\omega)\Gamma_j^\mathrm{B}g_{-r}^{\mathrm{R}}(x_2,x_1;\omega)\nonumber\\
&&+\Gamma_i^\mathrm{B}g_{r}^{\mathrm{R}}(x_i,x_j;\omega)\Gamma_i^\mathrm{B}g_{-r}^{\mathrm{R}}(x_i,x_i;\omega)[1-\Gamma_{\bar j}^\mathrm{F}g_{-r}^{\mathrm{R}}(x_{\bar j},x_{\bar j};\omega)]\bigg]~.\nonumber\\
\end{eqnarray}
The above formulas contain all the information necessary to calculate the LDOS of any chiral wire in the presence of one or two impurities as a function of energy and position. It is in fact a generalization of the solutions obtained in Ref.~\onlinecite{song03} for non interacting systems to take into account the effects of interactions. An important improvement with respect to Ref.~\onlinecite{song03} is the fact that Eqs.~(\ref{eq_dyson}), (\ref{Gxx1}), and (\ref{Gxx2}) are chirality resolved in order to include appropriately the Coulomb interactions.

In the following sections, we use these solutions to calculate the dependence of the density of states with energy and position for an infinite Luttinger liquid with different interaction strengths and impurity potentials, first for a QW with a single impurity (Sec. IV) and second, in the presence of two impurities (Sec. V).


\section{Results for an infinite spinless Luttinger liquid with a single impurity}\label{sec_single}

In this section, we consider a QW with a single impurity located at position $x_1=0$. We can thus take $\Gamma_{2}^{\mathrm{F,B}}=0$, and Eq.~(\ref{eq_dyson}) simplifies to
\begin{eqnarray}
&&G_{r,r'}^\mathrm{R}(x,x';\omega)= g_{r}^\mathrm{R}\left(x,x';\omega\right)\delta_{r,r'}+g_{r}^{\mathrm{R}}\left(x,x_1;\omega\right)\nonumber\\
&&\times\left[\Gamma_{1}^\mathrm{B} G_{-r,r'}^{\mathrm{R}}\left(x_1,x';\omega\right)
+\Gamma_{1}^\mathrm{F} G_{r,r'}^{\mathrm{R}}\left(x_1,x';\omega\right)\right]~.
\end{eqnarray}
The details of solving the above Dyson equation are presented in Appendix~\ref{solution_single_impurity}. We obtain
\begin{eqnarray}\label{Grr_single}
&&G_{r,r}^{\mathrm{R}}\left(x_1,x';\omega\right)=g_r^{\mathrm{R}}(x_1,x';\omega)\left[1-\Gamma^\mathrm{F}_1g_{-r}^{\mathrm{R}}(x_1,x_1;\omega)\right]\nonumber\\
&&\times\Big[1-\Gamma^\mathrm{F}_1\left[g_{r}^{\mathrm{R}}(x_1,x_1;\omega)+g_{-r}^{\mathrm{R}}(x_1,x_1;\omega)\right]\nonumber\\
&&+g_{r}^{\mathrm{R}}(x_1,x_1;\omega)\left[(\Gamma_1^\mathrm{F})^2-(\Gamma_1^\mathrm{B})^2\right]g_{-r}^{\mathrm{R}}(x_1,x_1;\omega)\Big]^{-1}~,\nonumber\\
\end{eqnarray}
and
\begin{eqnarray}\label{Gmrr_single}
&&G_{-r,r}^{\mathrm{R}}(x_1,x';\omega)=\frac{g_{-r}^{\mathrm{R}}\left(x_1,x_1;\omega\right)\Gamma_{1}^\mathrm{B}G_{r,r}^{\mathrm{R}}(x_1,x';\omega)}{1-\Gamma^\mathrm{F}_1g_{-r}^{\mathrm{R}}(x_1,x_1;\omega)}~.\nonumber\\
\end{eqnarray}
This allows us to determine fully the LDOS. 

\begin{figure}[H]
\begin{center}
\includegraphics[width=8.4cm]{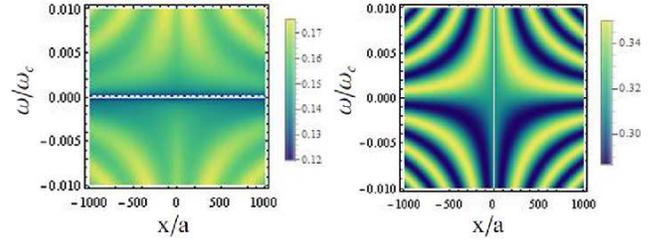}
\caption{LDOS (left graph) in the presence of Coulomb interactions ($K=0.7$), and (right graph) for a non interacting wire ($K=1$) as a function of position (horizontal axis) and energy (vertical axis) for a weak impurity $\Gamma_{1}^{\mathrm{F,B}}=0.1\hbar\omega_c$. We take $k_F=0$.}
\label{figure_SI_1}
\end{center} 
\end{figure}

\begin{figure}[H]
\begin{center}
\includegraphics[width=8.4cm]{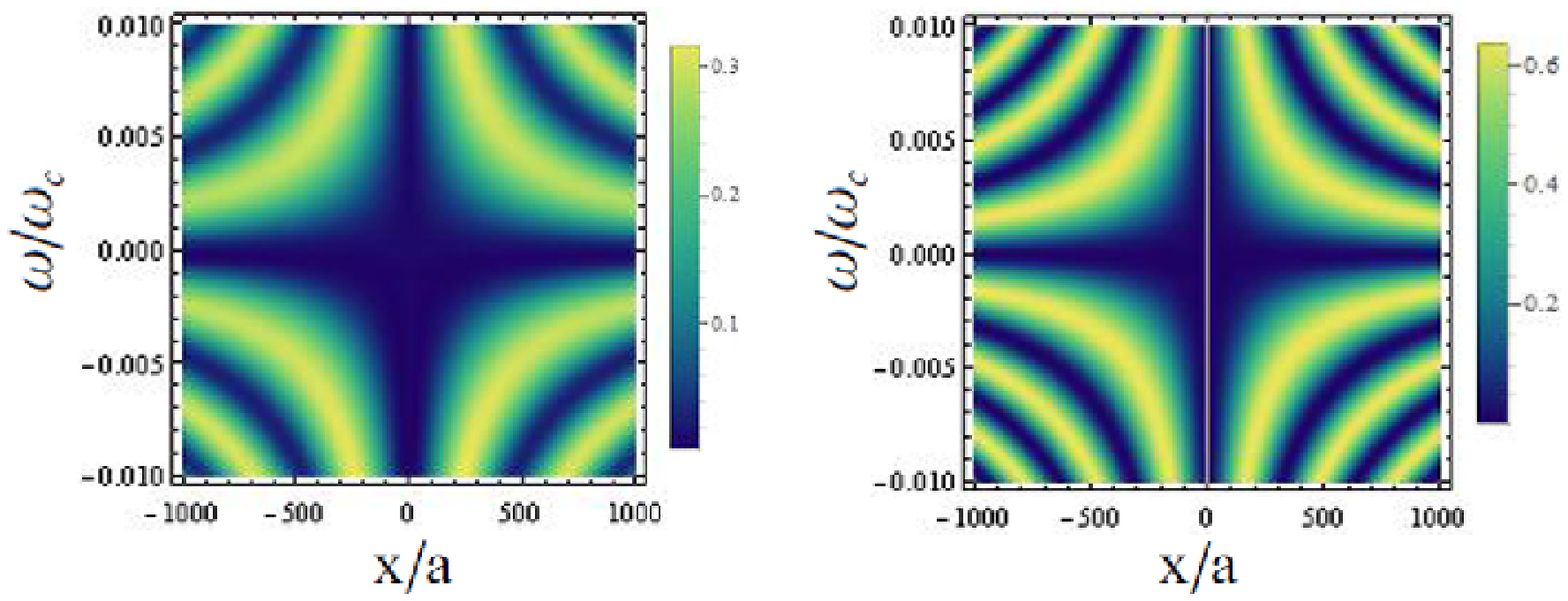}
\caption{The same as Fig.~\ref{figure_SI_1} for a strong impurity $\Gamma_{1}^{\mathrm{F,B}}=10\hbar\omega_c$.}
\label{figure_SI_2}
\end{center} 
\end{figure}

\begin{figure}[H]
\begin{center}
\includegraphics[width=4.2cm]{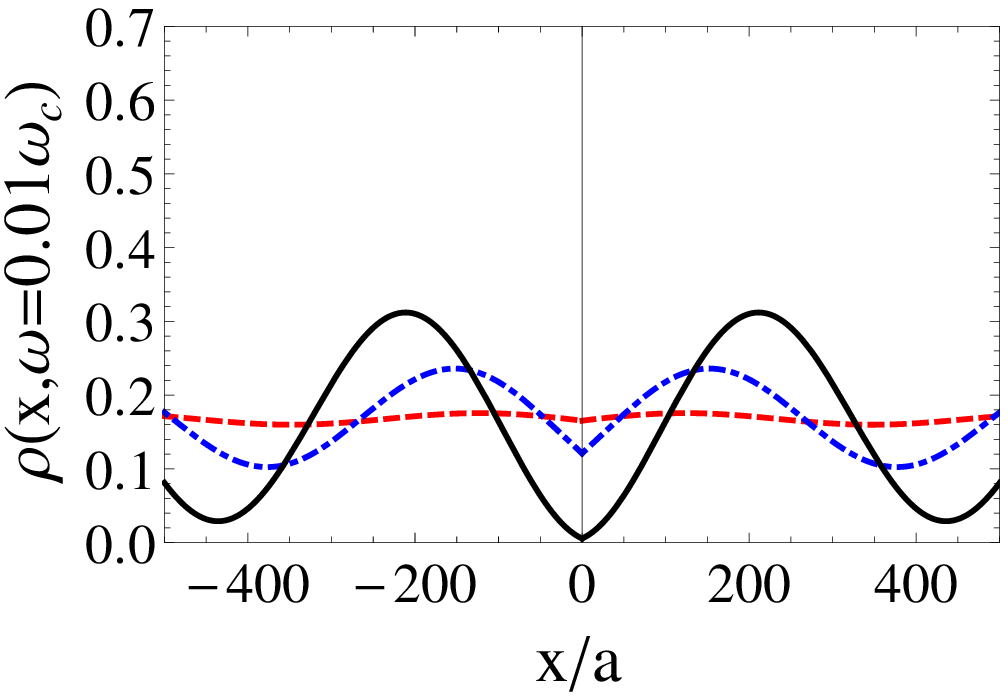}
\includegraphics[width=4.2cm]{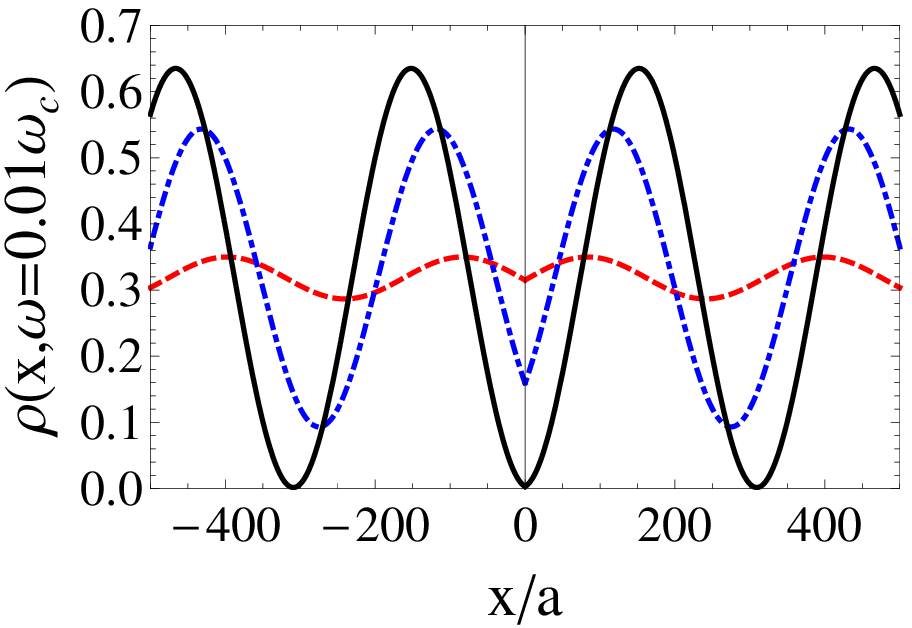}
\caption{LDOS in the presence of Coulomb interactions, $K=0.7$ (left graph) and for a non-interacting wire, $K=1$ (right graph) at $\omega=0.01\omega_c$, and for $\Gamma_{1}^{\mathrm{F,B}}=0.1\hbar\omega_c$ (red dashed lines), $\Gamma_{1}^{\mathrm{F,B}}=\hbar\omega_c$ (blue dashed-dotted lines), and $\Gamma_{1}^{\mathrm{F,B}}=10\hbar\omega_c$ (black solid lines). We take $k_F=0$.}
\label{figure_SI_3}
\end{center} 
\end{figure}

In Figs.~\ref{figure_SI_1} and \ref{figure_SI_2}, we show the profiles of the LDOS for increasing impurity potential in the presence and in the absence of Coulomb interactions. While the LDOS is asymmetric in energy at weak-impurity potential (Fig.~\ref{figure_SI_1}), it becomes symmetrical for the strong-impurity potential (Fig.~\ref{figure_SI_2}). The effect of the impurity is to introduce spatial oscillations whose amplitude increases with the impurity potential. In the presence of interactions,  the period of these Friedel oscillations is modified, and the value of the LDOS is reduced. As shown in Fig.~\ref{figure_SI_3}, the amplitude of oscillations and the density of states at the impurity position are both reduced for $K=0.7$ (left graph) in comparison to $K=1$ (right graph). 

\begin{figure}[H]
\begin{center}
\includegraphics[width=4.2cm]{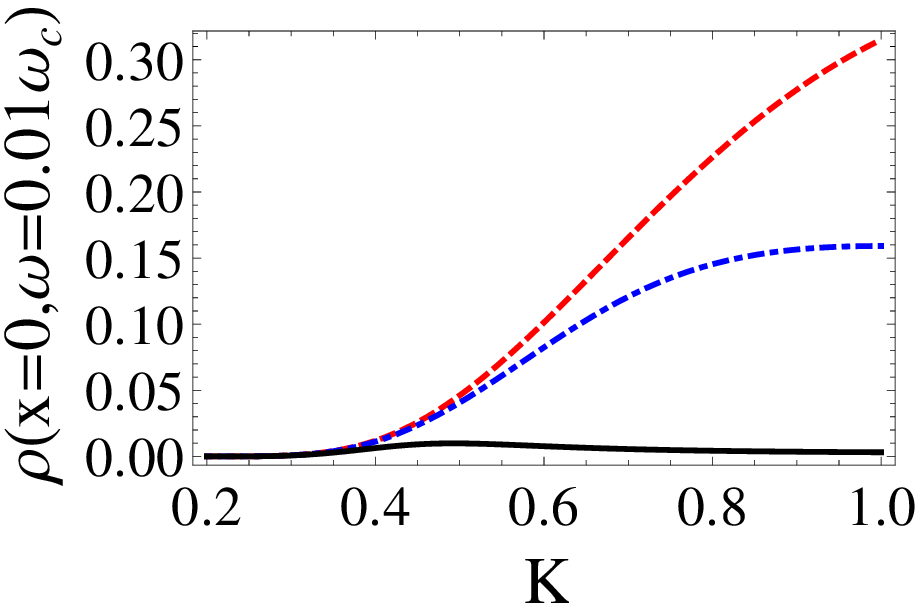}
\includegraphics[width=4.2cm]{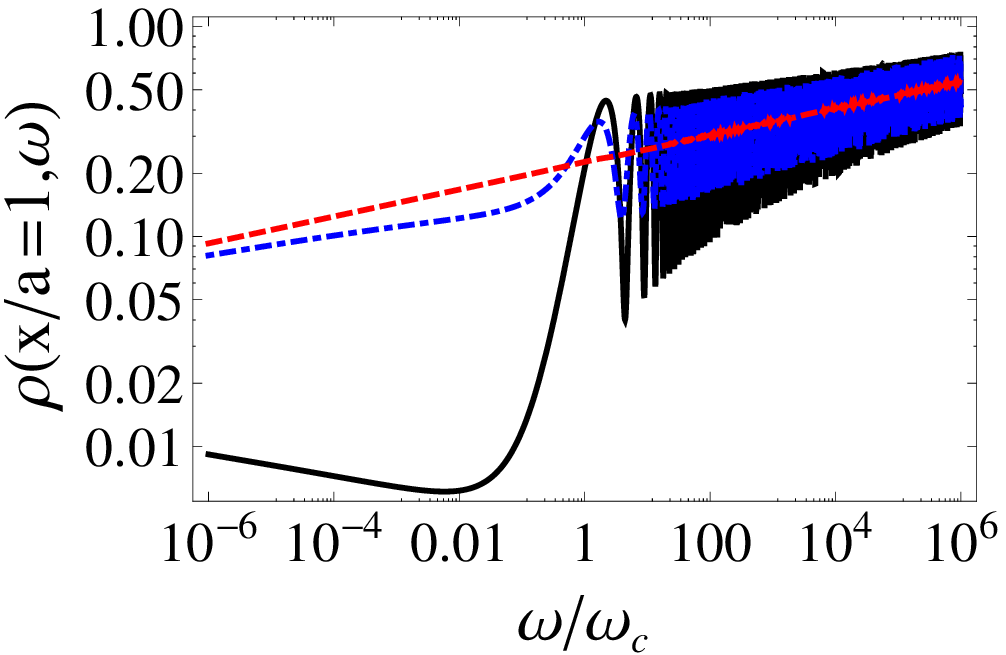}
\caption{LDOS (left graph) at the impurity position ($x=0$) as a function of $K$, at $\omega=0.01\omega_c$, and (right graph) close to the impurity position ($x/a=1$) as a function of energy at $K=0.7$. The right graph is plotted in a logarithm scale. On both graphs, we have $\Gamma_{1}^{\mathrm{F,B}}=0.01\hbar\omega_c$ (red dashed lines), $\Gamma_{1}^{\mathrm{F,B}}=\hbar\omega_c$ (blue dashed-dotted lines), and $\Gamma_{1}^{\mathrm{F,B}}=10\hbar\omega_c$ (black full lines). We take $k_F=0$.}
\label{figure_SI_4}
\end{center} 
\end{figure}

The reduction of the LDOS at the impurity position (here $x_1=0$) when increasing the strength of the interactions is observed for all values of the impurity potential (see the left graph in Fig.~\ref{figure_SI_4}). However, the LDOS is drastically reduced for the largest impurity potentials for all values of $K$ (black full line); this is because a large impurity effectively cuts the wire into two disconnected pieces.

In the right graph of Fig.~\ref{figure_SI_4} is shown the LDOS as a function of energy (on a logarithmic scale) for a position close to the impurity. For a weak impurity (red dashed line), the LDOS exhibits a power-law dependence with energy:
\begin{eqnarray}
\rho_0(\omega)=\frac{|\omega|^{(K+K^{-1}-2)/2}}{\pi{\bf \Gamma}\left(\frac{K+K^{-1}}{2}\right)}~,
\end{eqnarray}
which is just the density of states of a clean interacting wire. Here, $\bf{\Gamma}$ is the gamma function. When the impurity potential increases, the LDOS deviates from this power-law at small energy but converges and oscillates around this power law behavior when the energy increases, as expected (see the dashed-dotted blue line and the black line). This behavior is in full agreement with the results obtained in Refs.~\onlinecite{grishin04} and \onlinecite{eggert96}. However, within our approach (and its limitations), we are not able to recover the expected  behavior of the LDOS at low energy/distance and strong impurity potential, i.e., a power law behavior of $|\omega|^{K^{-1}-1}$ characteristic to injecting an electron into the end of a semi-infinite wire. As detailed in Appendix~\ref{approximation}, this is due to the fact that we neglect the terms mixing the Coulomb interactions and the impurity potential in the Dyson equation. This approximation is justified when $\Gamma^{\mathrm{B,F}}_1$ and $V_0$ are both weak. For strong $\Gamma^{\mathrm{B,F}}_1$, we need an addition assumption which is $|(x-x_1)\omega|\gg v_F$. This is the reason why our approach fails at low energy and strong impurity potentials. At high energy/large distances our results are valid. The range of validity of our approach depends on the impurity strength, the bulk power law being recovered for $|x-x_1|\gtrsim \hbar v/\Gamma^{\mathrm{B,F}}_1$ and $|\hbar\omega|\gtrsim\Gamma^{\mathrm{B,F}}_1$, i.e., when $|(x-x_1)\omega|\gtrsim v$, in agreement with Ref.~\onlinecite{grishin04}.

\section{Results for an infinite spinless Luttinger liquid with two impurities}

In this section, we turn our interest to a QW with two impurities located at positions $x_{1,2}=\pm L/2$.\\

\subsection{Density of states as a function of position and energy}\label{sec_profile}

\begin{figure}[H]
\begin{center}
\includegraphics[width=8.4cm]{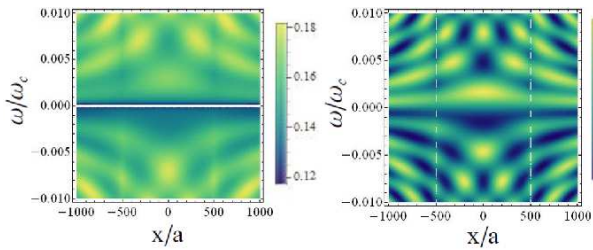}
\caption{LDOS (left graph) in the presence of Coulomb interactions ($K=0.7$), and (right graph) for a non-interacting wire ($K=1$) as a function of position (horizontal axis) and energy (vertical axis) for two symmetrical weak impurities $\Gamma_{1,2}^{\mathrm{F,B}}=0.1\hbar\omega_c$. The wire length is $L/a=1000$, and we take $k_F=0$.}
\label{figure2}
\end{center} 
\end{figure}

\begin{figure}[H]
\begin{center}
\includegraphics[width=8.4cm]{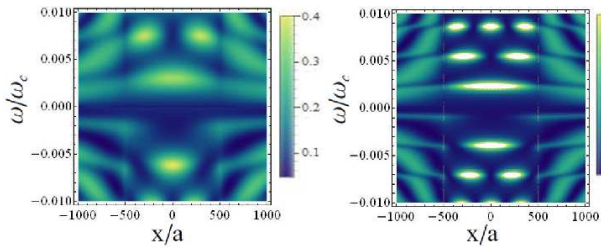}
\caption{The same as Fig.~\ref{figure2} for intermediate impurities $\Gamma_{1,2}^{\mathrm{F,B}}=\hbar\omega_c$.}
\label{figure3}
\end{center} 
\end{figure}

\begin{figure}[H]
\begin{center}
\includegraphics[width=8.4cm]{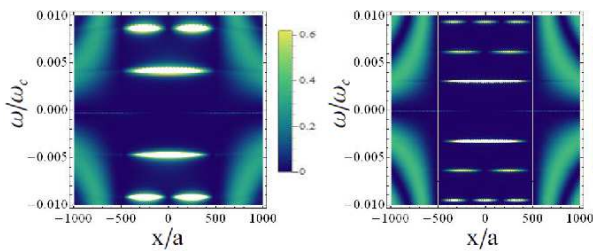}
\caption{The same as Fig.~\ref{figure2} for strong impurities $\Gamma_{1,2}^{\mathrm{F,B}}=10\hbar\omega_c$.}
\label{figure4}
\end{center} 
\end{figure}

We focus first on the analysis of the two-dimensional profiles of the LDOS as a function of position and energy.
In Figs.~\ref{figure2}, \ref{figure3}, and \ref{figure4}, we plot the profile of the density of states for increasing impurity potentials.

For small-impurity potentials (see Fig.~\ref{figure2}), we remark that the LDOS is odd in energy whereas it becomes even for strong-impurity potentials (see Fig.~\ref{figure4}), as it was the case for a single impurity. In the intermediate regime, the profile is neither odd, nor even (see Fig.~\ref{figure3}). Moreover, we note the evolution of the profiles from the weak-impurity Fabry-Perot regime to the strong-impurity, localized, Coulomb-blockade regime.\cite{bockrath99,rubio00,liang02,kim07} As previously shown for non interacting systems, in the Fabry-Perot weak-impurity regime the effect of the impurities reduces mostly to small sinusoidal oscillations of the LDOS with energy.\cite{song03} In the Coulomb-blockade regime, the strong-impurity potentials make the system evolve towards an isolated finite-size wire, whose LDOS is characterized by discrete energy levels, the distance between these levels being determined by the ratio between the Fermi velocity and the length of the central part of the wire, i.e., $\pi v_F/L$. As it can be seen from the left panels of Figs.~\ref{figure2}--\ref{figure4}, these characteristics persist in the presence of interactions\cite{tans97,bockrath97} except the distance between the discrete energy levels which becomes $\pi v/L$, with $v=v_F/K$.


The main differences between the profiles of the LDOS in the interacting and non interacting regimes consist in a modification of the amplitude and periodicity of the oscillations observed in the weak-impurity regime, as well as in a modification of the height, width, and periodicity of the peaks observed in the strong-impurity regime. Small features corresponding to a power-law reduction of the LDOS close to zero energy in the interacting limit are also present, although they are not very visible in the two-dimensional profiles. To study these points quantitatively, in what follows we study separately the dependence of the LDOS on  energy for a fixed position, as well as the dependence of the LDOS on the position for a given energy.

\subsection{Density of states as a function of energy}

\begin{figure}[H]
\begin{center}
\includegraphics[width=4.2cm]{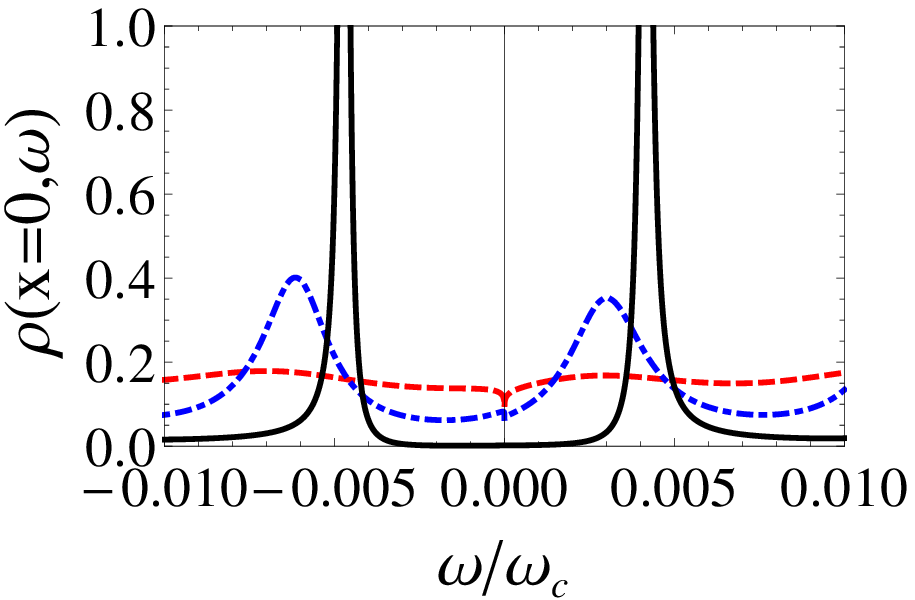}
\includegraphics[width=4.2cm]{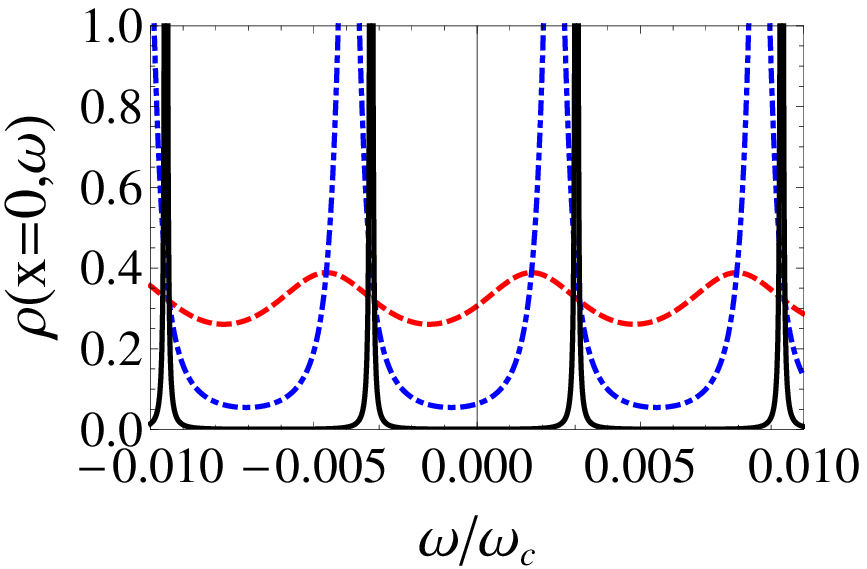}
\includegraphics[width=4.2cm]{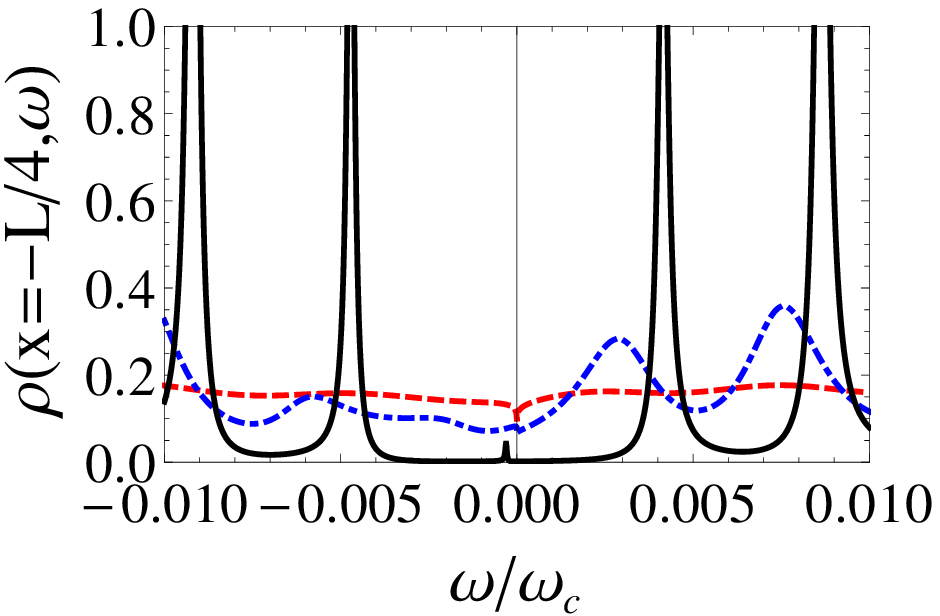}
\includegraphics[width=4.2cm]{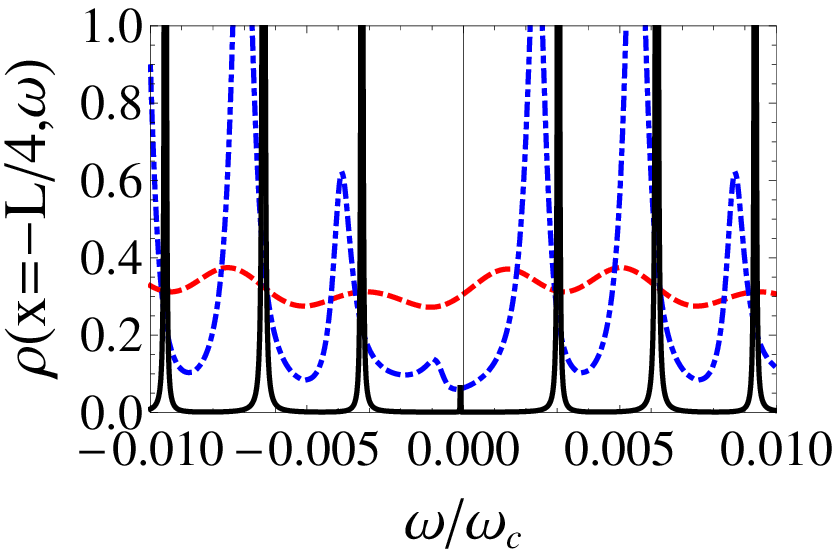}
\caption{LDOS (left graphs) in the presence of Coulomb interactions ($K=0.7$) and for a non interacting wire ($K=1$) (right graphs) as a function of energy at two different positions, $x=0$ (upper graphs) and $x=-L/4$ (bottom graphs), for $\Gamma_{1,2}^{\mathrm{F,B}}=0.1\hbar\omega_c$ (red dashed lines), $\Gamma_{1,2}^{\mathrm{F,B}}=\hbar\omega_c$ (blue dashed-dotted lines), and $\Gamma_{1,2}^{\mathrm{F,B}}=10\hbar\omega_c$ (black solid lines). We take $L/a=1000$ and $k_F=0$.}
\label{figure5}
\end{center} 
\end{figure}

Figure~\ref{figure5} shows the density of states as a function of energy for two different positions: $x=0$ (center of the wire) and $x=-L/4$ (halfway between the center of the wire and left impurity). For the interacting wire, in the weak impurity regime (see dashed red lines), as expected for a Luttinger liquid, a power-law reduction of the LDOS can be observed close to $\omega=0$. In the strong impurity regime (see solid black lines), as mentioned in the previous section, the central part of the wire is quasi-isolated and its spectrum resembles that of a finite size wire of length $L$ which is characterized by discrete peaks with energies of $n\pi v/L$, with $n$ being an integer. The height and width of the peaks depend on the coupling with the leads\cite{fisher97,bockrath97} which explains the sharpening of the peaks with increasing the strength of the impurity potentials. Note that depending on the position $x$, some peaks may not appear in the spectrum. Indeed, as it can be seen also from Fig.~\ref{figure4}, for $x=-L/4$, all the peaks are visible (see the solid black lines in the bottom graphs of Fig.~\ref{figure5}), 
whereas for $x=0$, only one peak out of two, corresponding to $\omega=(2n+1)\pi v/L$, is visible (see the solid black lines in the upper graphs of Fig.~\ref{figure5}).  This is due to the fact that we have a double periodicity: a first one with energy whose period is $n\pi v/L$ and a second with position whose period is related to the first one through $\pi v/\omega$, as Fig.~\ref{figure4} clearly shows.

Also, as already mentioned in Secs.~\ref{sec_single} and \ref{sec_profile}, the parity with respect to energy changes with impurity strength both for interacting and non interacting wires. To get some insight into these properties, we have performed a perturbative expansion of Eq.~(\ref{eq_dyson}) for $K=1$. In the weak-impurity regime, we find that the LDOS inside the wire can be written as
\begin{eqnarray}\label{asym_exp}
\Delta\rho(x,\omega)=\rho_0\sum_{i=1,2}\Gamma_i^{B}\sin\left(\frac{2\omega|x-x_i|}{v_F}\right)~,
\end{eqnarray}
with $\Delta \rho(x,\omega)= \rho(x,\omega)-\rho_0$, where $\rho_0$ is the density of states of the clean QW which takes a constant value for a non-interacting wire. This explains the odd parity of the LDOS described by the dashed red line in the upper right graph in Fig.~\ref{figure5}. We note that in this regime the impurity contribution to the LDOS is dominated by the backward-scattering terms. The forward scattering does not play any role since the amplitudes $\Gamma_1^\mathrm{F}$ and $\Gamma_2^\mathrm{F}$ drop out from the asymptotic expression of Eq.~(\ref{asym_exp}). This is not the case in the strong-impurity regime, for which we have found (not shown here) that the forward-scattering terms are essential to recover the Coulomb-blockade regime and cannot be neglected.

\begin{figure}[H]
\begin{center}
\includegraphics[width=4cm]{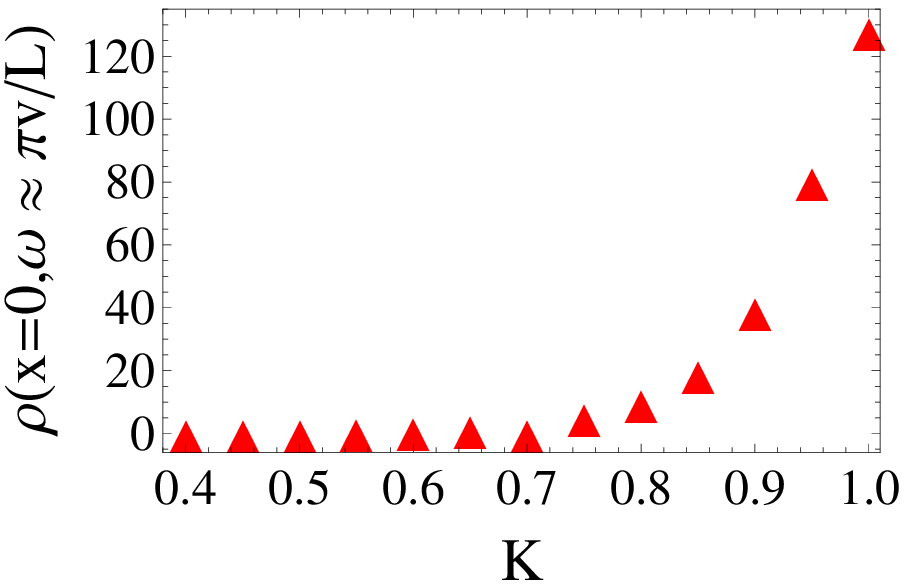}
\includegraphics[width=4.2cm]{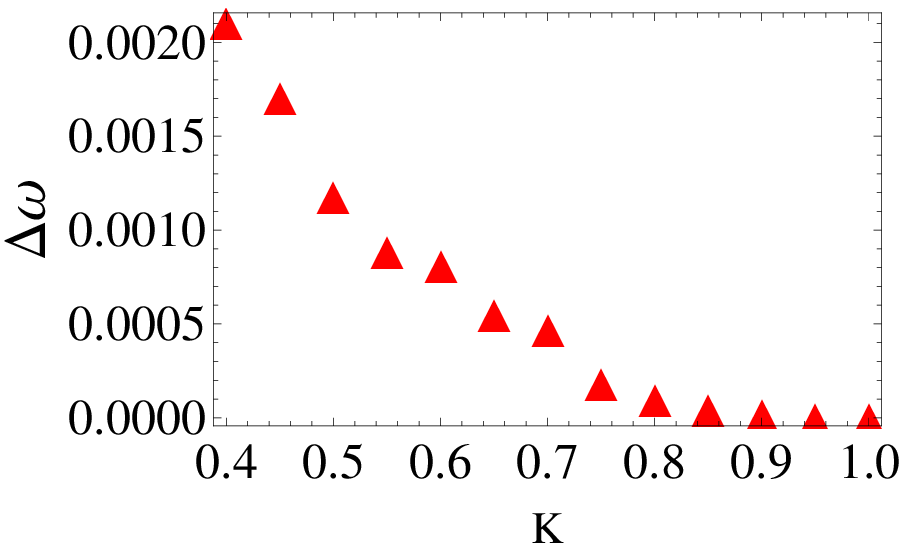}
\caption{Evolution of the peak amplitude (left graph) and peak width (right graph) as a function of $K$, for $\omega\approx\pi v/L$, $x=0$, and $\Gamma_{1,2}^{\mathrm{F,B}}=10\hbar\omega_c$. We take $L/a=1000$ and $k_F=0$.}
\label{figure6}
\end{center} 
\end{figure}

In order to understand how the effect of the Coulomb interactions affects the formation of the peaks in the LDOS,  in Fig.~\ref{figure6} we plot the height and the width  of the $\pi v/L$ peak as a function of $K$. When $K$ decreases (i.e., when Coulomb interactions increase), the peak broadens and its height is reduced. Moreover, the peak disappears completely for $K\lesssim 0.4$ and is replaced by an oscillating behavior of the LDOS. This result is quite intriguing as it would seem to indicate that, in what concerns the formation of the resonant levels, increasing the interactions effectively renormalizes the impurity strength to a smaller value, opposite to what would be intuitively expected from classical Luttinger liquid arguments. Note that both the distance between the peaks, and their positions are affected by the interaction strength.

\subsection{Density of states as a function of position}

\begin{figure}[H]
\begin{center}
\includegraphics[width=4.2cm]{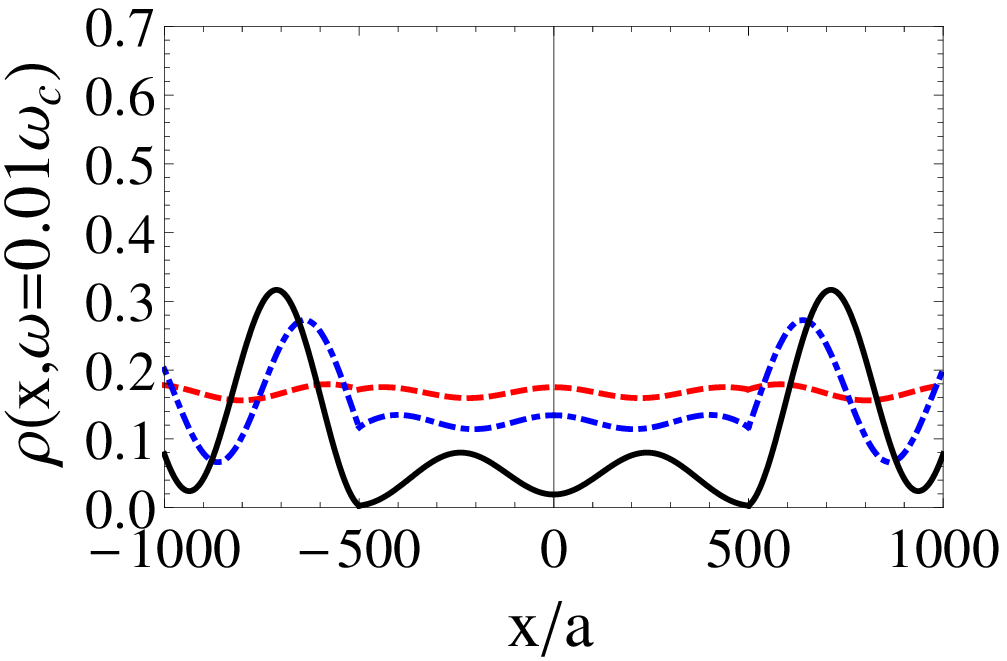}
\includegraphics[width=4.2cm]{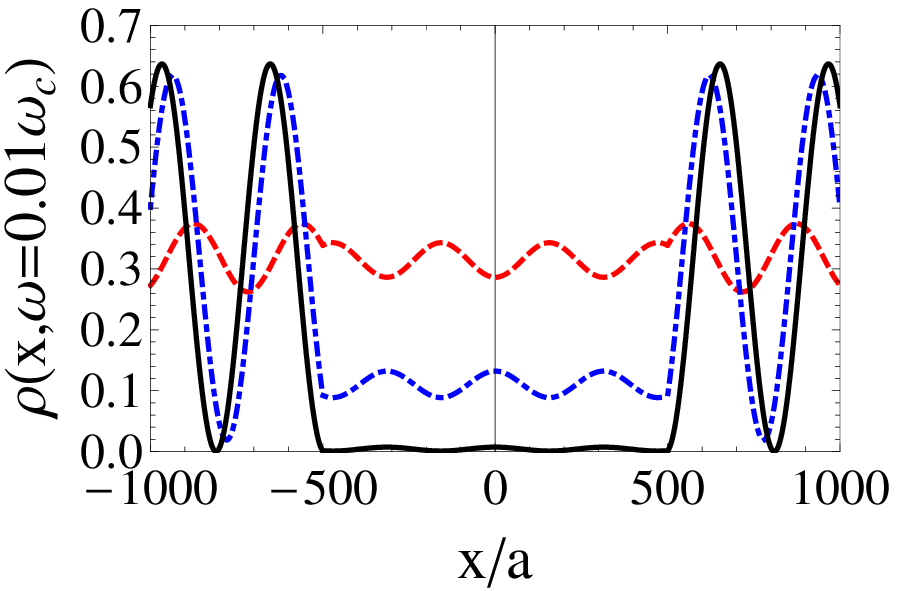}
\includegraphics[width=4cm]{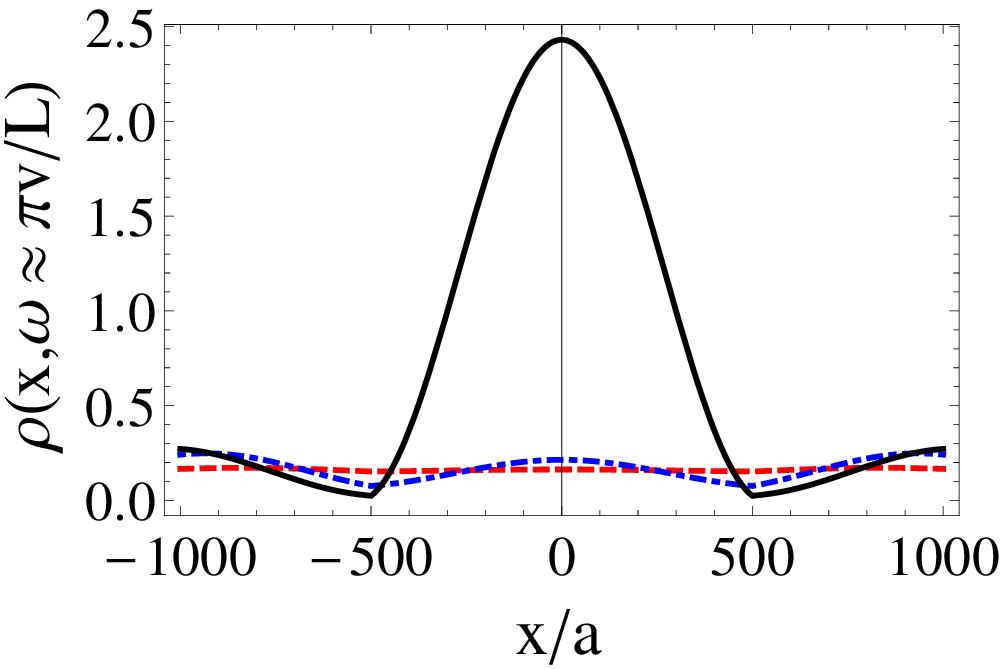}
\includegraphics[width=4.3cm]{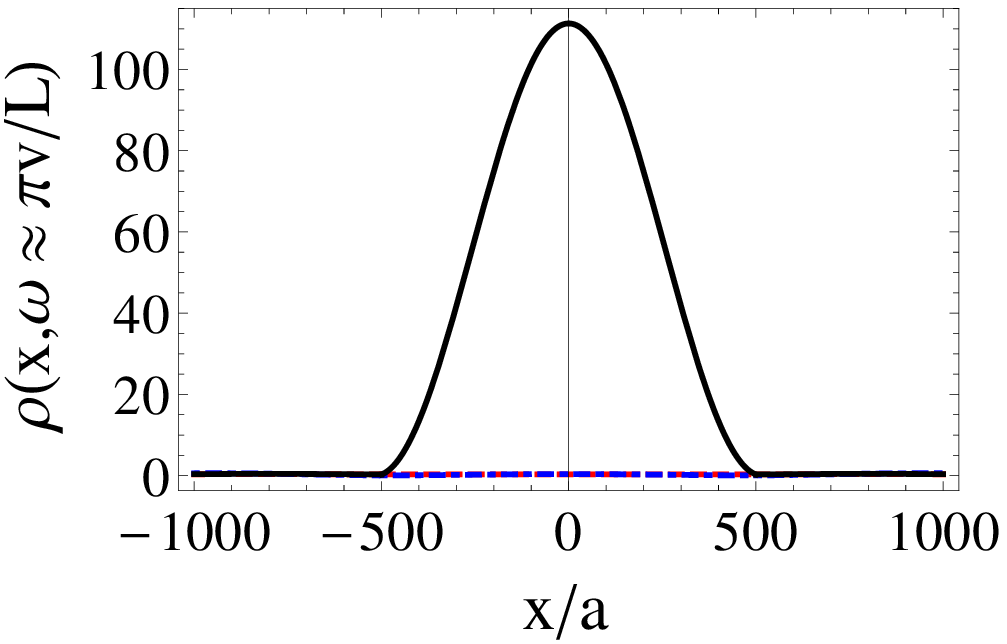}
\caption{LDOS for $K=0.7$ (left graph) and $K=1$ (right graph) as a function of position, for $\Gamma_{1,2}^{\mathrm{F,B}}=0.1\hbar\omega_c$ (red dashed lines), $\Gamma_{1,2}^{\mathrm{F,B}}=\hbar\omega_c$ (blue dash-dotted lines), and $\Gamma_{1,2}^{\mathrm{F,B}}=10\hbar\omega_c$  (black solid lines). The energy is taken to be $\omega=0.01\omega_c$ (upper graphs) and $\omega\approx \pi v/L$ (bottom graphs). We take $L/a=1000$ and $k_F=0$.}
\label{figure7}
\end{center} 
\end{figure}

We focus now on the dependence of the LDOS on the position for a given energy (see Fig.~\ref{figure7}). In the absence of interactions and in the strong impurity regime, the LDOS is uniformly zero in the central part of the wire for all energies that do not correspond to the formation of a peak (compare the solid black lines in the right upper and bottom graphs), as expected (see Ref.~\onlinecite{sablikov00}). In the presence of interactions, this reduction is less apparent (see the solid black lines in the left upper and bottom graphs) due to a competition between the Coulomb interactions and the oscillatory behavior related to the presence of impurities. However, even in the presence of Coulomb interactions, the LDOS in the strong impurity regime is zero at the impurity positions $x=x_{1,2}=\pm L/2$, since in this regime the wire is effectively disconnected from the leads (see the solid black lines in Fig.~\ref{figure7}).

\begin{figure}[t]
\begin{center}
\includegraphics[width=4.2cm]{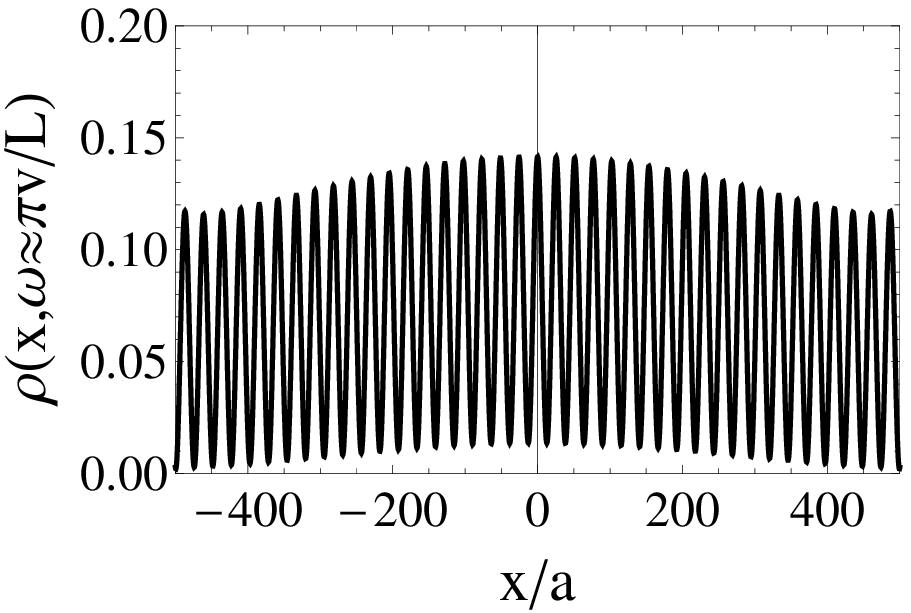}
\includegraphics[width=4.2cm]{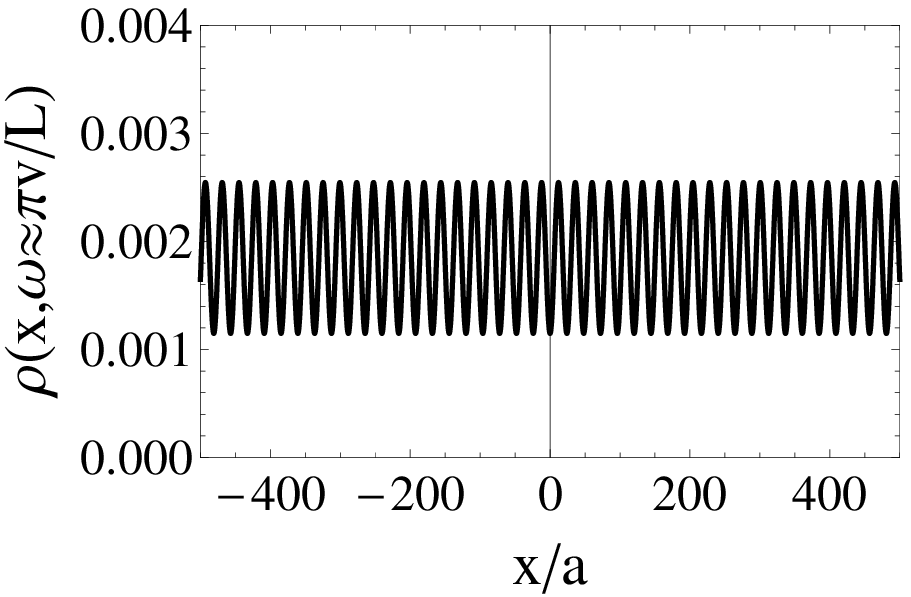}
\includegraphics[width=4.2cm]{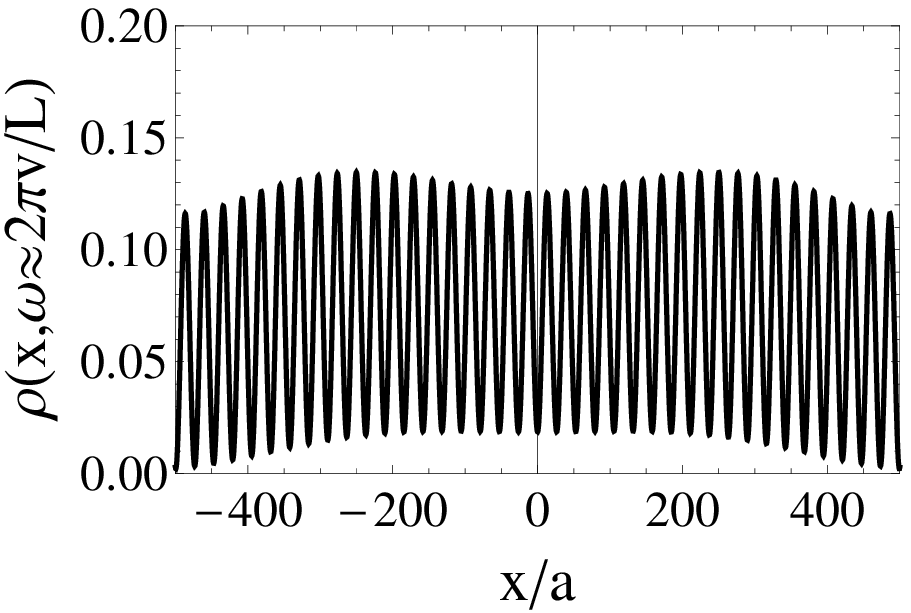}
\includegraphics[width=4.2cm]{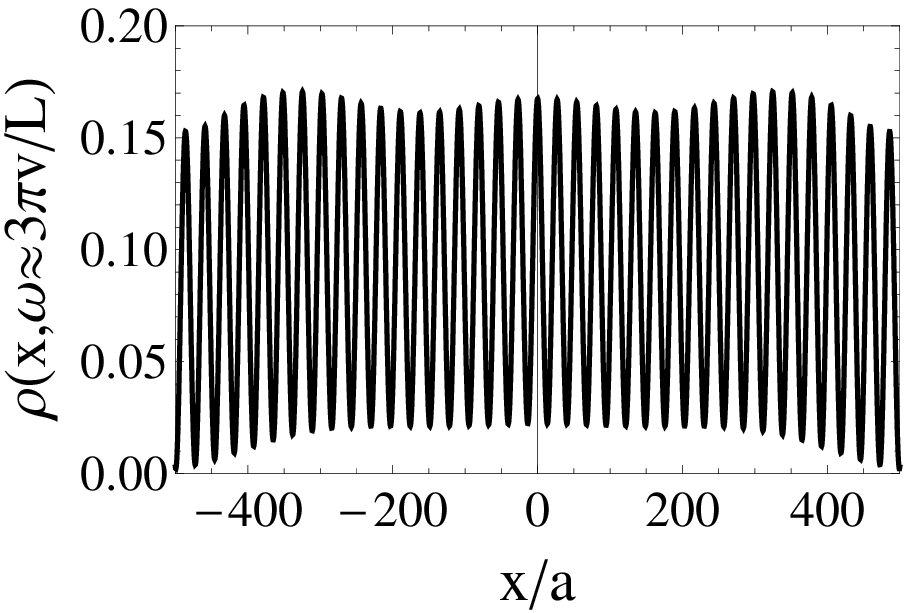}
\caption{LDOS for $x\in[-L/2,L/2]$, i.e., between the two impurities, in the presence of Coulomb interactions, $K=0.7$ (upper left graph and bottom graphs) and for a non-interacting wire, $K=1$ (upper right graph) at energy close to multiples of $\pi v/L$: either $\omega\approx \pi v/L$, $\omega\approx 2\pi v/L$ or $\omega\approx 3\pi v/L$ (see the legend of the vertical axis).  We take $k_F=40\pi/L$, $\Gamma_{1,2}^{\mathrm{F,B}}=10\hbar\omega_c$, and $L/a=1000$.}
\label{figure8}
\end{center} 
\end{figure}

Finally, we are interested to compare our results with those of Ref.~\onlinecite{anfuso03}. For this, we consider the dependence of the LDOS as a function of position at $k_F\ne 0$. In Fig.~\ref{figure8}, we plot the LDOS as a function of position at $\omega\approx \pi v/L$, and $k_F=40\pi/L$ for an interacting wire $K=0.7$ (upper left graph) and for a non-interacting wire $K=1$ (upper right graph). In both cases, the LDOS exhibits an oscillatory behavior whose period is $\pi/k_F$, but in the presence of Coulomb interactions, an extra-modulation appears and adds an envelope to the fast $\pi/k_F$ oscillations, in agreement with Ref.~\onlinecite{anfuso03}.  Close to the impurities, the amplitude of oscillations does not decrease, contrary to what is obtained in Ref.~\onlinecite{anfuso03}. This discrepancy comes, as outlined above, from exiting the limits of the regime of validity of our approach when we reach the impurity positions. However, recovering the same extra-modulation of the LDOS oscillations as those obtained in Ref.~\onlinecite{anfuso03} using a completely different technique, gives us an additional confirmation of the validity of our method at high energies and large distances with respect to the impurities. In Fig.~\ref{figure8} we plot also the LDOS in the presence of Coulomb interaction for $\omega\approx 2\pi v/L$ (bottom left graph) and for $\omega\approx 3\pi v/L$ (bottom right graph). It shows clearly that the period of the extra-modulation is equal to $\pi v/\omega$.



\section{Conclusion}
We have developed an approach based on the Dyson equations which has allowed us to study the LDOS of an infinite interacting QW with two impurities of arbitrary strength. For an infinite homogeneous interacting wire with a single impurity, we have calculated the form of the Friedel oscillations as well as the dependence of the LDOS with energy. We have found that for weak impurities, as well as for strong impurities at high energies/large distances, our approach recovers the expected Luttinger liquid power-law dependence; however, it breaks down for strong impurities at low energy/small distance. 

We have applied this approach to study the transition from the weak-impurity regime to the strong-impurity regime in a wire with two impurities, focusing in particular on the regime of large distances and energies. We  have found that the main effect of interactions is to reduce the amplitude of the Fabry-Perot oscillations in the weak-impurity limit, as well as of the Coulomb-blockade peaks in the strong-impurity limit. In addition, the interactions affect the periodicity of the oscillations and the distance between the peaks. Moreover, we see that strong interactions also reduce to zero the LDOS on the impurity sites. Also, at non zero $k_F$ and for strong impurities, our results are consistent with those obtained in Ref.~\onlinecite{anfuso03} for the LDOS of a Luttinger liquid in a box, in particular, we recover an extra-modulation of the LDOS oscillations in the presence of interactions, which is absent in the non-interacting system. This gives an extra confirmation that our approach is valid in that regime. 


Our work provides an important first step in constructing a non perturbative approach to understand the interplay between interactions and arbitrary size impurities in one-dimensional systems. Our approach is very general and can be easily applied to more realistic systems such as an inhomogeneous wire made of a central interacting region and two semi-infinite non interacting leads. To model such a system, one has to consider a spatial inhomogeneity in the interaction strength and use the Dyson equation method presented here, the fermionic Green's function for this system having been calculated in closed form.\cite{zamoum14} Our approach can also be generalized to take into account other factors such as the electronic spin which will allow us to characterize a more realistic system, such as carbon nanotube, and make connection with experiments.

Further improvements of this work would be to include long-range Coulomb interactions and to take into account the terms mixing Coulomb interactions and impurities potentials in the Dyson equation in order to fix the low-energy discrepancies between our results and those predicted by the perturbation theory in the strong-impurity limit.
\\
\\
{\bf Acknowledgements} 

We would like to thank Julia Meyer, Nicholas Sedlmayr, and Pascal Simon  for interesting discussions. The work of C.B. and M.G. is supported by the ERC Starting Independent Researcher Grant No. NANOGRAPHENE 256965.


\appendix

\section{Dyson equation derivation}\label{dyson}

To establish the Dyson equation for the retarded Green's function defined in Eq.~(\ref{G_R}), we evaluate $\partial_t\psi_r(x,t)=i[H;\psi_r(x,t)]/\hbar$, where $[a;b]$ denotes the commutator and $H$ is given by Eqs.~(\ref{H0})--(\ref{HF}). We obtain
\begin{eqnarray}
&&\partial_t\psi_{r}(x,t)=rv_F\partial_x\psi_{r}(x,t)\nonumber\\
&&-\frac{i}{\hbar}\sum_{i=1,2}\left[\lambda_{i}^\mathrm{B}(x) \psi_{-r}(x,t)
+\lambda_{i}^\mathrm{F}(x) \psi_{r}(x,t)\right]\nonumber\\
&&-\frac{i}{2\hbar}\sum_{r_1,r_2,r_3}\int_{-\infty}^\infty dx''V(x,x'')\nonumber\\
&&\times \Big\{\psi_{r_1}^\dag(x'',t)\psi_{r_2}(x'',t);\psi_{r_3}(x,t)\Big\}~.
\end{eqnarray}
Notice that the last term, which corresponds to the Coulomb interactions contribution, does not depend on $r$ since the interactions act similarly on both chiralities. After multiplying by the operator $\psi_{r'}^{\dagger}(x',t')$, we get:
\begin{eqnarray}\label{eq}
&&\partial_t\psi_{r}(x,t)\psi_{r'}^{\dagger}(x',t')=rv_F\partial_x\psi_{r}(x,t)\psi_{r'}^{\dagger}(x',t')\nonumber\\
&&-\frac{i}{\hbar}\sum_{i=1,2}\lambda_{i}^\mathrm{B}(x) \psi_{-r}(x,t)\psi_{r'}^{\dagger}(x',t')\nonumber\\
&&-\frac{i}{\hbar}\sum_{i=1,2}\lambda_{i}^\mathrm{F}(x) \psi_{r}(x,t)\psi_{r'}^{\dagger}(x',t')\nonumber\\
&&-\frac{i}{2\hbar}\sum_{r_1,r_2,r_3}\int_{-\infty}^\infty dx''V(x,x'')\nonumber\\
&&\times\bigg[\psi_{r_1}^\dag(x'',t)\psi_{r_2}(x'',t)\psi_{r_3}(x,t)\psi_{r'}^\dag(x',t')\nonumber\\
&&+\psi_{r_3}(x,t)\psi_{r_1}^\dag(x'',t)\psi_{r_2}(x'',t)\psi_{r'}^\dag(x',t')\bigg]~.
\end{eqnarray}
Because of Coulomb interactions,  Eq.~(\ref{eq}) contains products of four operators (see the two last lines of the above expression) which lead after averaging to the appearance of terms proportional to the two-particle Green's functions in the Dyson equation. In the presence of such terms, solving the Dyson equation is challenging and goes beyond the scope of this paper. Here, we make rather the assumption that the Coulomb potential $V(x,x'')$ decreases strongly when the distance $x-x''$ increases, and we keep only the dominant part in the integrals over $x''$, i.e., the $x\approx x''$ contribution, thus:
\begin{eqnarray}\label{der_op_0}
&&\partial_t\psi_{r}(x,t)\psi_{r'}^{\dagger}(x',t')=rv_F\partial_x\psi_{r}(x,t)\psi_{r'}^{\dagger}(x',t')\nonumber\\
&&-\frac{i}{\hbar}\sum_{i=1,2}\lambda_{i}^\mathrm{B}(x) \psi_{-r}(x,t)\psi_{r'}^{\dagger}(x',t')\nonumber\\
&&-\frac{i}{\hbar}\sum_{i=1,2}\lambda_{i}^\mathrm{F}(x) \psi_{r}(x,t)\psi_{r'}^{\dagger}(x',t')\nonumber\\
&&-\frac{i}{2\hbar}\sum_{r_1,r_2,r_3}V(x,x)\nonumber\\
&&\times\bigg[\psi_{r_1}^\dag(x,t)\psi_{r_2}(x,t)\psi_{r_3}(x,t)\psi_{r'}^\dag(x',t')\nonumber\\
&&+\underbrace{\psi_{r_3}(x,t)\psi_{r_1}^\dag(x,t)}_{=\delta_{r_3,r_1}-\psi_{r_1}^\dag(x,t)\psi_{r_3}(x,t)}\psi_{r_2}(x,t)\psi_{r'}^\dag(x',t')\bigg]~,
\end{eqnarray}
which reduces to:
\begin{eqnarray}\label{der_op_1}
&&\partial_t\psi_{r}(x,t)\psi_{r'}^{\dagger}(x',t')=rv_F\partial_x\psi_{r}(x,t)\psi_{r'}^{\dagger}(x',t')\nonumber\\
&&-\frac{i}{\hbar}\sum_{i=1,2}\lambda_{i}^\mathrm{B}(x) \psi_{-r}(x,t)\psi_{r'}^{\dagger}(x',t')\nonumber\\
&&-\frac{i}{\hbar}\sum_{i=1,2}\lambda_{i}^\mathrm{F}(x) \psi_{r}(x,t)\psi_{r'}^{\dagger}(x',t')\nonumber\\
&&-\frac{i}{\hbar}V_0\sum_{r_1}\psi_{r_1}(x,t)\psi_{r'}^\dag(x',t')~,
\end{eqnarray}
since compensations operate between the four operators terms of the fifth and sixth lines of Eq.~(\ref{der_op_0}) when one sums over $r_2$ and $r_3$. The only remaining contribution linked to Coulomb interactions is the one given by the fourth line in Eq.~(\ref{der_op_1}), where we have introduced $V_0=V(x,x)$. We can show in a similar manner that
\begin{eqnarray}\label{der_op_2}
&&\psi_{r'}^{\dagger}(x',t')\partial_t\psi_{r}(x,t)=rv_F\psi_{r'}^{\dagger}(x',t')\partial_x\psi_{r}(x,t)\nonumber\\
&&-\frac{i}{\hbar}\sum_{i=1,2}\lambda_{i}^\mathrm{B}(x) \psi_{r'}^{\dagger}(x',t')\psi_{-r}(x,t)\nonumber\\
&&-\frac{i}{\hbar}\sum_{i=1,2}\lambda_{i}^\mathrm{F}(x) \psi_{r'}^{\dagger}(x',t')\psi_{r}(x,t)\nonumber\\
&&-\frac{i}{\hbar}V_0\sum_{r_1}\psi_{r'}^{\dagger}(x',t')\psi_{r_1}(x,t)~.
\end{eqnarray}
Taking the time derivative of Eq.~(\ref{G_R}),
\begin{eqnarray}\label{der_green}
&&i\partial_tG_{r,r'}^\mathrm{R}(x,x';t,t')=\delta_{r,r'} \delta(x-x') \delta(t-t')\nonumber\\
&&+ \Theta(t-t') \langle \{ \partial_t\psi_{r}(x,t);\psi_{r'}^{\dagger}(x',t') \} \rangle~,
\end{eqnarray}
and substituting Eqs.~(\ref{der_op_1}) and (\ref{der_op_2}) in Eq.~(\ref{der_green}), we obtain
\begin{eqnarray}\label{diff_eq}
&&\hbar \left(i \partial_{t}-irv_F\partial_x\right)G_{r,r'}^\mathrm{R}(x,x';t,t')=\nonumber\\
&&\delta_{r,r'}\delta(x-x') \delta(t-t')+V_0\sum_{r_1}G_{r_1,r'}^\mathrm{R}(x,x';t,t')\nonumber\\
&&+\sum_{i=1,2} \Big[\lambda_{i}^\mathrm{B}(x)G_{-r,r'}^\mathrm{R}(x,x';t,t') \nonumber \\
&&+\lambda_{i}^\mathrm{F}(x)G_{r,r'}^\mathrm{R}(x,x';t,t')\Big]~.
\end{eqnarray}
Starting from this result, we can deduce immediately the equation of motion for the Green's function $g_{r}^\mathrm{0,R}$ in the absence of both impurities and Coulomb interactions, i.e., associated to the Hamiltonian $H_0$:
\begin{eqnarray}\label{diff_eq_g0}
&&\hbar \left(i \partial_{t}-irv_F\partial_x\right)g_{r}^\mathrm{0,R}(x,x';t,t')=\delta(x-x') \delta(t-t')~.\nonumber\\
\end{eqnarray}

From now, in order to simplify the notations, we remove the position and time dependence in Eq.~(\ref{diff_eq}) and insert $g_r^0$ in the expression of  $G_{r,r'}$:
\begin{eqnarray}\label{exact}
(g^0_r)^{-1}G_{r,r'}&=&\delta_{r,r'}+V_0\sum_{r_1}G_{r_1,r'}\nonumber\\
&&+\sum_{i=1,2} \Big[\lambda_{i}^\mathrm{B}G_{-r,r'} +\lambda_{i}^\mathrm{F}G_{r,r'}\Big]~.
\end{eqnarray}
It leads to:
\begin{eqnarray}\label{Eq1}
&&\Big[(g^0_r)^{-1}-V_0\Big]G_{r,r}=1+V_0G_{-r,r}\nonumber\\
&&+\sum_{i=1,2} \Big[\lambda_{i}^\mathrm{B}G_{-r,r} +\lambda_{i}^\mathrm{F}G_{r,r}\Big]~,
\end{eqnarray}
and,
\begin{eqnarray}\label{Eq2}
&&\Big[(g^0_{-r})^{-1}-V_0\Big]G_{-r,r}=V_0G_{r,r}\nonumber\\
&&+\sum_{i=1,2} \Big[\lambda_{i}^\mathrm{B}G_{r,r} +\lambda_{i}^\mathrm{F}G_{-r,r}\Big]~,
\end{eqnarray}
Inserting Eq.~(\ref{Eq2}) in ~(\ref{Eq1}), we get:
\begin{eqnarray}\label{Eq3}
&&\Big[(g^0_r)^{-1}-V_0-V_0[(g^0_{-r})^{-1}-V_0]^{-1}V_0\Big]G_{r,r}=\nonumber\\
&&1+\sum_{i=1,2} \Big[\lambda_{i}^\mathrm{B}G_{-r,r} +\lambda_{i}^\mathrm{F}G_{r,r}\Big]\nonumber\\
&&+V_0g^0_{-r}\Big[1-V_0g^0_{-r}\Big]^{-1}\sum_{i=1,2} \Big[\lambda_{i}^\mathrm{B}G_{r,r} +\lambda_{i}^\mathrm{F}G_{-r,r}\Big]~.\nonumber\\
\end{eqnarray}
Assuming that the terms involving the product of the impurity potential and the interaction potential in the third line of Eq.~(\ref{Eq3}) are negligible (see Appendix \ref{approximation} for justification), we end up with
\begin{eqnarray}\label{Eq4}
\Big[(g^0_r)^{-1}-V_0-V_0[(g^0_{-r})^{-1}-V_0]^{-1}V_0\Big]G_{r,r}=\nonumber\\
1+\sum_{i=1,2} \Big[\lambda_{i}^\mathrm{B}G_{-r,r} +\lambda_{i}^\mathrm{F}G_{r,r}\Big]~.
\end{eqnarray}

After some algebra, we find without further assumptions that the factor in front of $G_{r,r}$ in Eq.~(\ref{Eq4}) corresponds to the inverse of the Green's function of an interacting clean QW, i.e., associated to $H_0+H_\mathrm{int}$, namely,
\begin{eqnarray}
g_r&=&\Big[(g^0_r)^{-1}-V_0-V_0[(g^0_{-r})^{-1}-V_0]^{-1}V_0\Big]^{-1}\nonumber\\
&=&[1-g^0_{-r}V_0]\underbrace{\Big[1-\sum_{r_1}g^0_{r_1}V_0\Big]^{-1}}_{=V_0^{-1}\Sigma_\mathrm{int}}g^0_r\nonumber\\
&=&V_0^{-1}\Sigma_\mathrm{int}g^0_r-g^0_{-r}\Sigma_\mathrm{int}g^0_r\nonumber\\
&=&V_0^{-1}\Big[V_0+V_0\sum_{r_1}g^0_{r_1}\Sigma_\mathrm{int}\Big]g^0_r-g^0_{-r}\Sigma_\mathrm{int}g^0_r\nonumber\\
&=&g^0_r+g^0_r\Sigma_\mathrm{int}g^0_r~,
\end{eqnarray}
where the self-energy associated to the Coulomb interactions is defined as
\begin{eqnarray}
\Sigma_\mathrm{int}&=&V_0+V_0\sum_{r_1=\pm}g^0_{r_1}\Sigma_\mathrm{int}\nonumber\\
&=&V_0+V_0\sum_{r_1=\pm}g^0_{r_1}V_0+V_0\sum_{r_1=\pm}g^0_{r_1}V_0\sum_{r_2=\pm}g^0_{r_2}V_0+\dots.\nonumber\\
\end{eqnarray}
The key point in this calculation is the fact that the self-energy contains a sum over the two chiralities. Finally, Eq.~(\ref{Eq4}) can be written as:
\begin{eqnarray}\label{Eq5}
G_{r,r}&=&g_r+\sum_{i=1,2} \Big[g_r\lambda_{i}^\mathrm{B}G_{-r,r} +g_r\lambda_{i}^\mathrm{F}G_{r,r}\Big]~.
\end{eqnarray}
Similarly, we can show using the same approximations that:
\begin{eqnarray}\label{Eq6}
G_{r,-r}&=&\sum_{i=1,2} \Big[g_r\lambda_{i}^\mathrm{B}G_{-r,-r} +g_r\lambda_{i}^\mathrm{F}G_{r,-r}\Big]~.
\end{eqnarray}
Notice that the Coulomb potential $V_0$ does not appear explicitly in Eqs.~(\ref{Eq5}) and (\ref{Eq6}) because it has been incorporated in the Green's function of the clean interacting wire, $g_{r}$.

Restoring the position and time dependences in Eqs.~(\ref{Eq5}) and (\ref{Eq6}), we have:

\begin{eqnarray}\label{dyson_eq}
G_{r,r'}^\mathrm{R}(x,x';t,t')&=&g_{r}^\mathrm{R}(x,x';t,t')\delta_{r,r'} \nonumber\\ 
&&+\sum_{i=1,2}\int dt''\int dx'' g_{r}^\mathrm{R}(x,x'';t,t'')  \nonumber \\
&&\times \left[\lambda_{i}^\mathrm{B}(x'')G_{-r,r'}^\mathrm{R}(x'',x';t'',t') \right.\nonumber \\
&&+\left.\lambda_{i}^\mathrm{F}(x'')G_{r,r'}^\mathrm{R}(x'',x';t'',t')\right]~,
\end{eqnarray}
which leads, after a Fourier transform, to Eq.~(\ref{eq_dyson}), when one assumes delta-function localized impurities, i.e., $\lambda_i^{\mathrm{F,B}}(x)=\Gamma_i^{\mathrm{F,B}}\delta(x-x_i)$ for $i=1,2$. We insist on the fact that this equation is valid when either both $V_0$ and $\lambda_i^{\mathrm{B,F}}$ are small in comparison to energy, or at high energy $\omega$ and large distance $x-x_i$ to the impurities when $\lambda_i^{\mathrm{B,F}}$ is large (see Appendix \ref{approximation} for the details).


\section{Domain of validity}\label{approximation}

In this appendix, we discuss the domain of validity of the approximations used in our work.

We first consider the regime where both $V_0$ and $\lambda_i^{\mathrm{B,F}}$ are small in comparison to the other energy of the problem, namely, $\hbar\omega$. In that case, we can expand Eq.~(\ref{exact}), which is exact, up to first-order terms in $V_0$ and $\lambda_i^{\mathrm{B,F}}$:
\begin{eqnarray}
G_{r,r}^{(1)}&=&g_{r}^0+g_{r}^0V_0g_{r}^0+\sum_{i=1,2}g_{r}^0\lambda_{i}^\mathrm{F}g_{r}^0~,\\
G_{r,-r}^{(1)}&=&g_{r}^0V_0g_{-r}^0+\sum_{i=1,2}g_{r}^0\lambda_{i}^\mathrm{B}g_{-r}^0~,
\end{eqnarray}
which coincide with the expansions of Eqs.~(\ref{Eq5}) and~(\ref{Eq6}) except concerning an additional term $g_{r}^0V_0g_{-r}^0$ absent in Eq.~(\ref{Eq6}) which, however, gives a negligible contribution to the LDOS since it is proportional to $V_0/\hbar\omega$. Our approach is thus not limited by any additional assumption when both $V_0$ and $\lambda_i^{\mathrm{B,F}}$ are small in comparison to $\hbar\omega$.

We now consider the regime where either $V_0$ or $\lambda_i^{\mathrm{B,F}}$ is large, in that case, we have to consider higher-order terms in the expansion with $V_0$ and $\lambda_i^{\mathrm{B,F}}$. Here, we restrict our discussion up to the second-order expansions of Eq.~(\ref{exact}):
\begin{eqnarray}\label{Grr_2order}
G_{r,r}^{(2)}&=&G_{r,r}^{(1)}+g_{r}^0V_0\sum_{r_1=\pm}g_{r_1}^0V_0g_{r}^0\nonumber\\
&+&\sum_{i=1,2}g_{r}^0\bigg[V_0g_{r}^0\lambda_{i}^\mathrm{F}+\lambda_{i}^\mathrm{F}g_{r}^0V_0\bigg] g_{r}^0\nonumber\\
&+&\sum_{i=1,2}g_{r}^0\bigg[V_0g_{-r}^0\lambda_{i}^\mathrm{B}+\lambda_{i}^\mathrm{B}g_{-r}^0V_0\bigg] g_{r}^0\nonumber\\
&+&\sum_{i,j=1,2}g_{r}^0\bigg[\lambda_{i}^\mathrm{B}g_{-r}^0\lambda_{j}^\mathrm{B}+\lambda_{i}^\mathrm{F}g_{r}^0\lambda_{j}^\mathrm{F}\bigg] g_{r}^0~,
\end{eqnarray}
and,
\begin{eqnarray}\label{Grmr_2order}
G_{r,-r}^{(2)}&=&G_{r,-r}^{(1)}+g_{r}^0V_0\sum_{r_1=\pm}g_{r_1}^0V_0g_{-r}^0\nonumber\\
&+&\sum_{i=1,2}g_{r}^0\bigg[V_0g_{-r}^0\lambda_{i}^\mathrm{F}+\lambda_{i}^\mathrm{F}g_{r}^0V_0\bigg] g_{-r}^0\nonumber\\
&+&\sum_{i=1,2}g_{r}^0\bigg[V_0g_{r}^0\lambda_{i}^\mathrm{B}+\lambda_{i}^\mathrm{B}g_{-r}^0V_0\bigg] g_{-r}^0\nonumber\\
&+&\sum_{i,j=1,2}g_{r}^0\bigg[\lambda_{i}^\mathrm{B}g_{-r}^0\lambda_{j}^\mathrm{F}+\lambda_{i}^\mathrm{F}g_{r}^0\lambda_{j}^\mathrm{B}\bigg] g_{-r}^0~.
\end{eqnarray}

Equations~(\ref{Grr_2order}) and~(\ref{Grmr_2order}) coincide with the expansions of Eqs.~(\ref{Eq5}) and~(\ref{Eq6}) except concerning the additional following terms: the terms $g_{r}^0V_0g_{-r}^0\lambda_{i}^\mathrm{B}g_{r}^0$ and $g_{r}^0\lambda_{i}^\mathrm{B}g_{-r}^0V_0g_{r}^0$ of the third line in Eq.~(\ref{Grr_2order}), and the terms $g_{r}^0V_0g_{-r}^0\lambda_{i}^\mathrm{F}g_{-r}^0$ and $g^0_r\lambda_i^\mathrm{F}g^0_rV_0g^0_{-r}$ of the second line in Eq.~(\ref{Grmr_2order}). These four terms are missing in Eqs.~(\ref{Eq5}) and~(\ref{Eq6}) that we have use in this work. The amplitude of these terms can be estimated using Eq.~(\ref{gnon}), remembering that we have to restore the position dependencies and to add an integration over position each time that we have a product of two Green's functions. Performing these integrations, we find that the missing terms are all of the order of $(V_0\Gamma_{i}^\mathrm{B,F}v_F/\omega)e^{2ir\omega(x-x_i)/v_F}$, which is small in comparison to the contributions to the LDOS that we have considered adequately, i.e., those of the third line of Eq.~(\ref{Grmr_2order}):
\begin{eqnarray}
 g_{r}^0V_0 g_{r}^0\lambda_{i}^\mathrm{B}g_{-r}^0&\propto& (x-x_i)V_0\Gamma_{i}^\mathrm{B}e^{2ir\omega(x-x_i)/v_F}~,\\
 g_{r}^0\lambda_{i}^\mathrm{B}g_{-r}^0V_0 g_{-r}^0&\propto& (x-x_i)V_0\Gamma_{i}^\mathrm{B}e^{2ir\omega(x-x_i)/v_F}~,
\end{eqnarray}
provided that $|(x-x_i)\omega|\gg v_F$. This argument can be reproduced at any order in the expansion with $V_0$ and $\lambda_i^{\mathrm{B,F}}$ since the terms we neglect are always of the form $g_{r}^0V_0g_{-r}^0$. Notice that we do not consider  in this comparison the terms $g_{r}^0V_0g_{r}^0\lambda_{i}^\mathrm{F}g_{r}^0$ and $g_{r}^0\lambda_{i}^\mathrm{F}g_{r}^0V_0g_{r}^0$ of the second line in Eq.~(\ref{Grr_2order}) since they do not contribute at all to the LDOS. Thus, at large $\lambda_i^{\mathrm{B,F}}$, our approach is valid provided that $|(x-x_i)\omega|\gg v_F$. Besides, we need to keep the assumption $V_0/\hbar\omega\ll 1$ in order to be allowed to neglect the contribution $g_{r}^0V_0g_{-r}^0$ in $G_{r,-r}^{(1)}$.


\section{Solution of the Dyson equation for two impurities}\label{solution}

By taking first, $x=x_1$, and second, $x=x_2$, in Eq.~(\ref{eq_dyson}) with $r=\pm r'$, we obtain a set of linear coupled equations (all the frequency arguments have been dropped in order to simplify the notations):
\begin{eqnarray}\label{C1}
G_{r,r}^{\mathrm{R}}(x_1,x')&=& g_{r}^\mathrm{R}\left(x_1,x'\right)\nonumber\\
&+&g_{r}^{\mathrm{R}}\left(x_1,x_1\right) \Gamma_{1}^\mathrm{B} G_{-r,r}^{\mathrm{R}}\left(x_1,x'\right)\nonumber\\
&+&g_{r}^{\mathrm{R}}\left(x_1,x_2\right) \Gamma_{2}^\mathrm{B} G_{-r,r}^{\mathrm{R}}\left(x_2,x'\right)\nonumber\\
&+&g_{r}^{\mathrm{R}}\left(x_1,x_1\right) \Gamma_{1}^\mathrm{F} G_{r,r}^{\mathrm{R}}\left(x_1,x'\right)\nonumber\\
&+&g_{r}^{\mathrm{R}}\left(x_1,x_2\right) \Gamma_{2}^\mathrm{F} G_{r,r}^{\mathrm{R}}\left(x_2,x'\right)~,
\end{eqnarray}
\begin{eqnarray}\label{C2}
G_{r,r}^{\mathrm{R}}(x_2,x')&=& g_{r}^\mathrm{R}\left(x_2,x'\right)\nonumber\\
&+&g_{r}^{\mathrm{R}}\left(x_2,x_1\right) \Gamma_{1}^\mathrm{B} G_{-r,r}^{\mathrm{R}}\left(x_1,x'\right)\nonumber\\
&+&g_{r}^{\mathrm{R}}\left(x_2,x_2\right) \Gamma_{2}^\mathrm{B} G_{-r,r}^{\mathrm{R}}\left(x_2,x'\right)\nonumber\\
&+&g_{r}^{\mathrm{R}}\left(x_2,x_1\right) \Gamma_{1}^\mathrm{F} G_{r,r}^{\mathrm{R}}\left(x_1,x'\right)\nonumber\\
&+&g_{r}^{\mathrm{R}}\left(x_2,x_2\right) \Gamma_{2}^\mathrm{F} G_{r,r}^{\mathrm{R}}\left(x_2,x'\right)~,
\end{eqnarray}
and,
\begin{eqnarray}
\label{C3}
G_{-r,r}^{\mathrm{R}}(x_1,x')&=&
g_{-r}^{\mathrm{R}}\left(x_1,x_1\right) \Gamma_{1}^\mathrm{B} G_{r,r}^{\mathrm{R}}\left(x_1,x'\right)\nonumber\\
&+&g_{-r}^{\mathrm{R}}\left(x_1,x_2\right) \Gamma_{2}^\mathrm{B} G_{r,r}^{\mathrm{R}}\left(x_2,x'\right)\nonumber\\
&+&g_{-r}^{\mathrm{R}}\left(x_1,x_1\right) \Gamma_{1}^\mathrm{F} G_{-r,r}^{\mathrm{R}}\left(x_1,x'\right)\nonumber\\
&+&g_{-r}^{\mathrm{R}}\left(x_1,x_2\right) \Gamma_{2}^\mathrm{F} G_{-r,r}^{\mathrm{R}}\left(x_2,x'\right)~,
\end{eqnarray}
\begin{eqnarray}\label{C4}
G_{-r,r}^{\mathrm{R}}(x_2,x')&=&
g_{-r}^{\mathrm{R}}\left(x_2,x_1\right) \Gamma_{1}^\mathrm{B} G_{r,r}^{\mathrm{R}}\left(x_1,x'\right)\nonumber\\
&+&g_{-r}^{\mathrm{R}}\left(x_2,x_2\right) \Gamma_{2}^\mathrm{B} G_{r,r}^{\mathrm{R}}\left(x_2,x'\right)\nonumber\\
&+&g_{-r}^{\mathrm{R}}\left(x_2,x_1\right) \Gamma_{1}^\mathrm{F} G_{-r,r}^{\mathrm{R}}\left(x_1,x'\right)\nonumber\\
&+&g_{-r}^{\mathrm{R}}\left(x_2,x_2\right) \Gamma_{2}^\mathrm{F} G_{-r,r}^{\mathrm{R}}\left(x_2,x'\right)~.
\end{eqnarray} 
From Eqs.~(\ref{C3}) and (\ref{C4}), we extract the expressions of $G_{-r,r}^{\mathrm{R}}$ and express them only in terms of the Green's functions $G_{r,r}^{\mathrm{R}}$. Thus, we get:
\begin{eqnarray}\label{C5}
G_{-r,r}^{\mathrm{R}}(x_1,x')&=&D^{-1}\sum_{j=1,2}g_{-r}^{\mathrm{R}}\left(x_1,x_j\right)\Gamma_{j}^\mathrm{B}G_{r,r}^{\mathrm{R}}(x_j,x')\nonumber\\
&&+D^{-1}\Gamma_1^\mathrm{B}\Gamma_2^\mathrm{F}\Big[g_{-r}^{\mathrm{R}}(x_1,x_2)g_{-r}^{\mathrm{R}}(x_2,x_1)\nonumber\\
&&-g_{-r}^{\mathrm{R}}(x_1,x_1)g_{-r}^{\mathrm{R}}(x_2,x_2)\Big]G_{r,r}^{\mathrm{R}}(x_1,x')~,\nonumber\\
\end{eqnarray}
and,
\begin{eqnarray}\label{C6}
G_{-r,r}^{\mathrm{R}}(x_2,x')&=&D^{-1}\sum_{j=1,2}g_{-r}^{\mathrm{R}}\left(x_2,x_j\right)\Gamma_{j}^\mathrm{B}G_{r,r}^{\mathrm{R}}(x_j,x')\nonumber\\
&&+D^{-1}\Gamma_2^\mathrm{B}\Gamma_{1}^\mathrm{F}\Big[g_{-r}^{\mathrm{R}}(x_2,x_1)g_{-r}^{\mathrm{R}}(x_1,x_2)\nonumber\\
&&-g_{-r}^{\mathrm{R}}(x_2,x_2)g_{-r}^{\mathrm{R}}(x_1,x_1)\Big]G_{r,r}^{\mathrm{R}}(x_2,x')~,\nonumber\\
\end{eqnarray}
where $D$ is defined by Eq.~(\ref{def_D}). Equations~(\ref{C5}) and (\ref{C6}) correspond to Eq.~(\ref{Gxx2}).
Next, substituting Eqs.~(\ref{C5}) and (\ref{C6}) in Eqs.~(\ref{C1}) and (\ref{C2}), we end up with a system of two linear equations whose solutions are:
\begin{eqnarray}\label{C7}
G_{r,r}^{\mathrm{R}}\left(x_1,x';\omega\right)&=&\frac{(1-\chi_r^{22})g_r^{\mathrm{R}}(x_1,x';\omega)+\chi_r^{12}g_r^{\mathrm{R}}(x_{2},x';\omega)}{(1-\chi_r^{11})(1-\chi_r^{22})-\chi_r^{12}\chi_r^{21}}~,\nonumber\\
\end{eqnarray}
and,
\begin{eqnarray}\label{C8}
G_{r,r}^{\mathrm{R}}\left(x_2,x';\omega\right)&=&\frac{(1-\chi_r^{11})g_r^{\mathrm{R}}(x_2,x';\omega)+\chi_r^{21}g_r^{\mathrm{R}}(x_{1},x';\omega)}{(1-\chi_r^{11})(1-\chi_r^{22})-\chi_r^{12}\chi_r^{21}}~,\nonumber\\
\end{eqnarray}
where $\chi_r^{ij}$ is defined by Eq.~(\ref{def_chi}). Equations~(\ref{C7}) and (\ref{C8}) correspond to Eq.~(\ref{Gxx1}).


\section{Solution of the Dyson equation for a single impurity}\label{solution_single_impurity}

We consider a single impurity located at position $x_1$. In that case, $\Gamma_{2}^\mathrm{B}=\Gamma_{2}^\mathrm{F}=0$ and Eq.~(\ref{dyson_eq}) simplifies. After a Fourier transform, we obtain:
\begin{eqnarray}
&&G_{r,r'}^\mathrm{R}(x,x';\omega)= g_{r}^\mathrm{R}\left(x,x';\omega\right)\delta_{r,r'}+g_{r}^{\mathrm{R}}\left(x,x_1;\omega\right)\nonumber\\
&&\times\left[\Gamma_{1}^\mathrm{B} G_{-r,r'}^{\mathrm{R}}\left(x_1,x';\omega\right)
+\Gamma_{1}^\mathrm{F} G_{r,r'}^{\mathrm{R}}\left(x_1,x';\omega\right)\right]~.
\end{eqnarray} 
We can extract the expressions of $G_{r,r}^{\mathrm{R}}\left(x_1,x';\omega\right)$ and $G_{-r,r}^{\mathrm{R}}\left(x_1,x';\omega\right)$ by solving a linear set of equations as done in Appendix \ref{solution}. We obtain:
\begin{eqnarray}\label{single_Grr}
G_{r,r}^{\mathrm{R}}\left(x_1,x';\omega\right)&=&\frac{g_r^{\mathrm{R}}(x_1,x';\omega)}{1-\chi_r^{11}}~,
\end{eqnarray}
and,
\begin{eqnarray}\label{single_Gmrr}
&&G_{-r,r}^{\mathrm{R}}(x_1,x';\omega)=\frac{g_{-r}^{\mathrm{R}}\left(x_1,x_1;\omega\right)\Gamma_{1}^\mathrm{B}G_{r,r}^{\mathrm{R}}(x_1,x';\omega)}{1-\Gamma_1^\mathrm{F}g_{-r}^{\mathrm{R}}\left(x_1,x_1;\omega\right)}~,\nonumber\\
\end{eqnarray} 
where $\chi_r^{11}$ for a single impurity reduces to:
\begin{eqnarray}\label{exp_chi}
\chi_r^{11}&=&\Gamma_1^Fg_r^{\mathrm{R}}(x_1,x_1;\omega)\nonumber\\
&&+\frac{g_r^{\mathrm{R}}(x_1,x_1;\omega)(\Gamma_1^B)^2g_{-r}^{\mathrm{R}}(x_1,x_1;\omega)}{1-\Gamma_1^\mathrm{F}g_{-r}^{\mathrm{R}}\left(x_1,x_1;\omega\right)}~.
\end{eqnarray}

Replacing Eq.~(\ref{exp_chi}) in Eqs.~(\ref{single_Grr}) and (\ref{single_Gmrr}), we end up with Eqs.~(\ref{Grr_single}) and (\ref{Gmrr_single}).


\end{document}